\documentclass{article}
\usepackage{authblk}

\usepackage{geometry}
\usepackage{hyperref}
\hypersetup{
    colorlinks,
    linkcolor={black},
    citecolor={black},
    urlcolor={black}
}
\usepackage{epstopdf, epsfig}

\usepackage{tabularx}
\usepackage{tikz}
\usepackage{bigints}
\usepackage{float}
\usepackage{comment}
\usepackage{caption}
\usepackage{subcaption}
\usepackage{soul}
\usepackage{color,graphicx,float}
\usepackage{graphicx}
\usepackage{soul}
\usepackage{bm}
\usepackage{array}
\usepackage{makecell}
\usepackage[round,authoryear]{natbib}
\usepackage{physics}
\usepackage{float}
\usepackage{todo}
\usepackage{enumitem}
\usepackage{multirow}
\usepackage{tabularx}
\usepackage{gensymb}
\usepackage{amsmath}
\usepackage{enumerate}
\usepackage{mathrsfs}
\usepackage{color,graphicx,float}
\usepackage{hyperref}
\hypersetup{colorlinks,linkcolor={blue}, citecolor={blue}, urlcolor={black}}

\newcommand{\red}[1]{\ifmmode\mathbf{\textcolor{red}{#1}}\else \textbf{\textcolor{red}{#1}}\fi}

\newcommand{\blue}[1]{\ifmmode{\textcolor{blue}{#1}}\else {\textcolor{blue}{#1}}\fi}
\usepackage{xcolor}

\def\F{\mathcal{F}}
\def\cl{\text{CL}}

\title{Droplet coalescence on a sloped fibre}

\author[1]{Souradip Chattopadhyay}
\author[2]{Leyun Feng}
\author[2]{Kyoo-Chul Park}
\author[1]{Hangjie Ji\thanks{Corresponding author: \href{mailto:hangjie_ji@ncsu.edu}{hangjie\_ji@ncsu.edu}}}

\affil[1]{Department of Mathematics, North Carolina State University, Raleigh, NC 27695, USA}

\affil[2]{Department of Mechanical Engineering, Northwestern University, Evanston, IL 60208, USA}

\date{}

\begin{document}

\maketitle 

\begin{abstract}
Transporting droplets along cylindrical surfaces is an emerging area of research with a wide range of practical applications such as fog collection and filtration. Recent experimental studies by \cite{feng2024short} have revealed a wealth of new droplet coalescence dynamics on a pre-wetted cylindrical fibre, illustrating the need for more advanced theory. In this paper, we study both the early-stage and late-stage dynamics of droplet coalescence on a sloped cylindrical fibre. In the early stage,  when the minimum thickness of the liquid bridge connecting two droplets is small, we propose a lubrication-based model to analyze the capillary-driven self-similar dynamics of the liquid bridge. Our theory predicts the bridge growth rate and demonstrates that asymmetry in the adjoining contact angles contributes to the migration of the merged droplet, showing good agreement with experimental observations.
In the late stage, as droplet merging progresses, inertia effects become significant and lead to rapid mass transfer. 
To capture this regime, we develop a weighted residual model for  liquid films flowing on a sloped pre-wetted cylinder, incorporating key physical effects such as gravity, low-to-moderate inertia, surface tension, and intermolecular forces. This model explains the experimentally observed directional self-propelled transport during droplet coalescence, where the larger droplet migrates toward the smaller one.
Numerical simulations based on this model further support these findings, showing good agreement between the predicted coalescence-induced droplet migration and experimental observations.
\end{abstract}


\section{Introduction}\label{sec:1}

Liquid droplet formation and coalescence on cylindrical substrates is a fundamental process in both natural phenomena (such as on spider webs) and engineering applications. In fog harvesting and mist elimination devices (\cite{jiang2019fog,brunazzi2000design,kowalski2022dynamics}), hundreds of cylindrical fibres form mesh structures that promote droplet formation and coalescence in high-humidity environments. The dynamics of traveling droplets coating a cylindrical fibre have been studied for applications such as particle and vapor capture (\cite{sadeghpour2019water,sadeghpour2021experimental}). 
Droplet-fibre systems have also been utilized in digital microfluidics (\cite{gilet2009digital}).
Having a better understanding of the mechanisms that drive the droplet dynamics on cylindrical surfaces is essential to optimize the design of these devices.

\par In a typical binary droplet merging event, two droplets meet and form a connecting liquid bridge. This process is dominated by localized surface tension at early times (\cite{eggers1999coalescence}) and can enter the inertial regime (\cite{sprittles2014dynamics}) as the liquid bridge grows according to scaling laws, followed by rapid mass transfer between the two droplets that eventually leads to droplet merging (\cite{ashgriz1990coalescence}). For a comprehensive review of droplet coalescence, readers are referred to \cite{eggers2024coalescence}.

\par Both early-stage and late-stage coalescence dynamics can strongly depend on the geometry of the substrate.
For instance, \cite{hernandez2012symmetric} investigated the symmetric and asymmetric coalescence of drops on a flat substrate, describing the bridge profile using similarity solutions of a one-dimensional lubrication equation.
\cite{lorenceau2004drops} experimentally showed that viscous droplets on a conical fibre spontaneously move toward the region of lower curvature.
Further experiments on the capillary-driven migration of droplets on conical fibres include the works of
\cite{li2013fastest} and \cite{fournier2021droplet}.
\cite{chan2020directional} developed a lubrication model along with matched asymptotics to study the directional spreading of a viscous droplet on a conical fibre. The formation of liquid films on conical fibres has been studied experimentally and numerically using the lubrication model by \cite{chan2021film}. \cite{pawar2019symmetric} proposed a model to describe the coalescence of two droplets on a substrate, considering both equal and unequal sizes. Using the lattice Boltzmann method, they demonstrated that when droplets are either perfect hemispheres or in a spreading state, the liquid bridge height follows the same scaling law for both cases. More recent studies \citep{dekker2022elasticity,kaneelil2025coalescence} have also focused on the impact of viscoelasticity on droplet coalescence.

\par Incorporating gravity into droplet-fibre systems can also lead to complex droplet morphologies and dynamics. 
\cite{christianto2022modeling} studied the gravity-driven dynamics of partially wetting droplets on fibres in both small and large Bond number regimes using the lattice Boltzmann method.
More recently,  \cite{bharti2023plateau} investigated the Rayleigh-Plateau instability of a thin viscous film on a soft fibre.
While most studies have focused on axisymmetric droplets on cylindrical and conical domains, the work of \cite{gupta2021effect} examined the effects of gravity on the shape of nearly-axisymmetric droplets.

\par Recent experimental studies (\cite{jiang2022coalescence, feng2024short}) have revealed that directional liquid transport on a cylindrical fibre can be achieved through coalescence-induced propulsion. The authors showed that a larger droplet can be pulled toward a smaller one in asymmetric droplet coalescence dynamics, even in the presence of gravity. 
\begin{figure}
    \centering
    \includegraphics[width=0.7\linewidth]{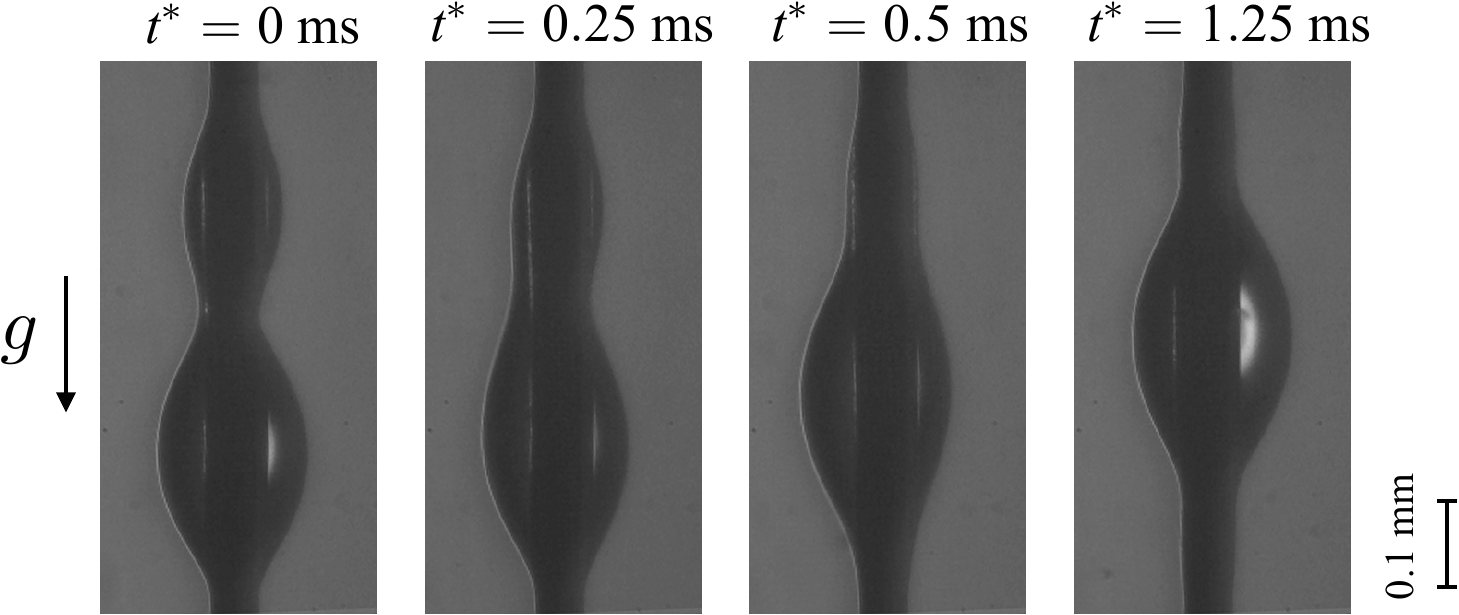}
    \caption{Coalescence of two water droplets on a pre-wetted vertical cylindrical fibre exhibiting upward movement of the larger droplet against gravity during the coalescence event. See the supplementary movie provided in movie1.mp4.
    }
    \label{fig:experiment_90}
\end{figure}
Figure \ref{fig:experiment_90} shows a binary water droplet coalescence experiment on a vertical cylindrical fibre of radius $R^*=0.025$ mm, where initially a smaller droplet is positioned above a larger droplet. A thin meniscus bridge is established between the two droplets at time $t^* = 0$, which then rapidly grows, pulling the larger droplet toward the smaller one, followed by the merging of two droplets at around $t^* = 0.5$ ms. The droplet coalescence triggers an upward movement of the merged droplet, as observed in the experimental profile at time $t^* = 1.25$ ms. This directional transport is fundamentally different from asymmetric droplet coalescence on a flat substrate, where the smaller droplet is typically drawn toward the larger one by the Laplace pressure difference, while the larger droplet resists motion because of its greater inertia.

\par In the work of \cite{feng2024short},
this unprecedented droplet transport dynamics was explained using a simple mass-spring-damper model. Their model represents the coalescing droplets as point masses connected by an effective spring and damper, together with assumptions on droplet-size-dependent viscous friction. While this reduced-order model captures the overall migration of the merged droplet, it does not resolve the underlying free-surface dynamics or the fluid transport during the coalescence process. In particular, the evolution of the liquid bridge during the early capillary-driven stage and the subsequent inertia-influenced mass transfer remain poorly understood.

\par In this work, we present a continuum hydrodynamics framework that systematically describes both stages of droplet coalescence on a sloped cylindrical fibre. For the early stage, we develop a lubrication model that predicts the asymptotically self-similar growth of the liquid bridge and quantifies the influence of asymmetric droplet coalescence on the directional motion. For the late stage, we derive a weighted-residual model that incorporates gravity, surface tension, low-to-moderate inertia, and intermolecular forces to capture the merging and migration dynamics. The proposed models show good agreement with experiments and provide further insights into the roles of inertia, friction, surface tension, unequal parent droplet sizes, and fibre inclination angles in the directional migration of the merged droplet.

\par The structure of the paper is as follows. The experimental setup for typical droplet coalescence is presented in section \ref{sec:experiment}. In section \ref{sec:model}, we present the model formulation for the droplet dynamics using the weighted residual approach. In section \ref{sec:early_coal},  we study the early stage of droplet coalescence driven by capillary forces using lubrication theory and compare the numerical results with experimental observations. In section \ref{sec:later_coal}, we discuss late-stage coalescence dynamics influenced by moderate inertia effects using the weighted residual model. Concluding remarks and a discussion of remaining open questions are presented in section \ref{sec:discussion}.

\section{Experiments}
\label{sec:experiment}
\begin{figure}
    \centering
\includegraphics[width=15cm]{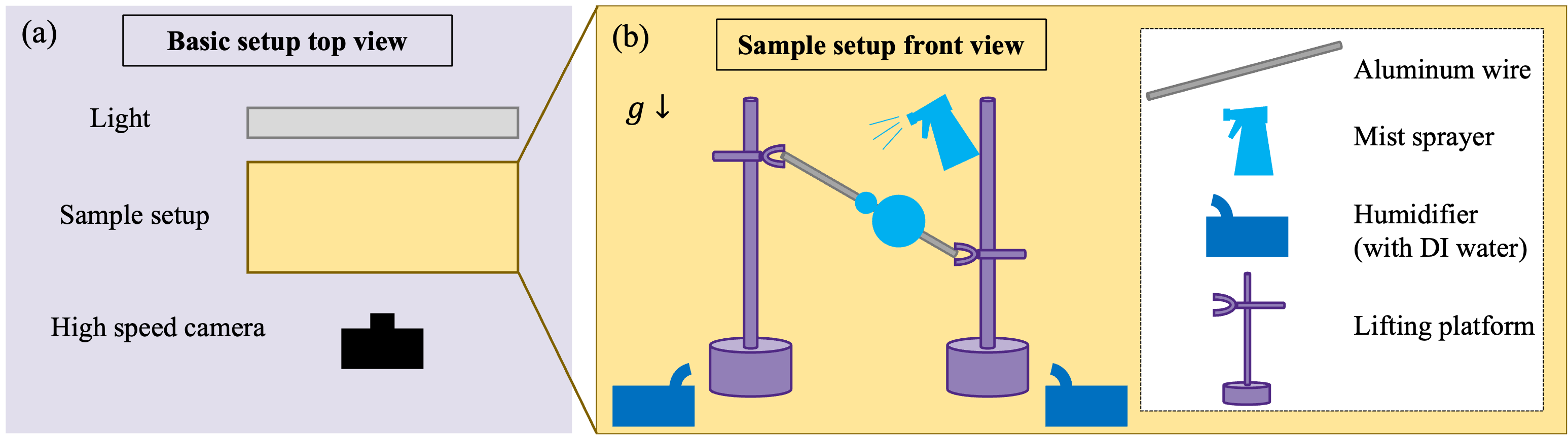}
    \caption{Experiment setup.}
    \label{fig:exp}
\end{figure}
\emph{Sample preparation.} Aluminum fibres were purchased from Alfa Aesar. The aluminum fibres were first cleaned in an ultrasonic cleaner (Branson) with a detergent (Alcojet) for 30 minutes and then rinsed with deionized (DI) water. Then, the cleaned fibres and sheets were placed in boiling DI water for $30$ minutes to produce nanostructures ($\gamma$-AlOOH), in a process known as boehmitization \citep{jiang2019fog,jiang2022coalescence}. The boehmitized surfaces are superhydrophilic. 
The working fluid is DI water with density $\rho = 998$ kg m$^{-3}$, kinematic viscosity $\nu = 1$ mm$^2$
 s$^{-1}$, and surface tension $\sigma = 72$ mN m$^{-1}$ at $20~^{\degree}$C. The experiments are performed using prepared aluminum fibres of diameter $0.05$ mm.
 
\emph{Measuring droplet motion.} Figure \ref{fig:exp} shows the experimental setup used in this study. Parent droplets were deposited on a pre-wetted superhydrophilic fibre by a sprayer (Houseables) and grown by additionally deposited airborne droplets generated from humidifiers (Pure enrichment). The droplet interaction and the resulting motion were captured by acquiring images at $20000$ fps using a high-speed camera (Phantom FASTCAM Mini AX200) and magnifying lens (Nikon CF PLAN $20X/0.35$ SLWD Microscope Objective). The height and length of each droplet before and after coalescence were determined with Tracker (a free video analysis and modeling tool built on the Open Source Physics (OSP) Java framework). 

\section{Model formulation}\label{sec:model}
\begin{figure}
    \centering
    \includegraphics[scale=0.4]{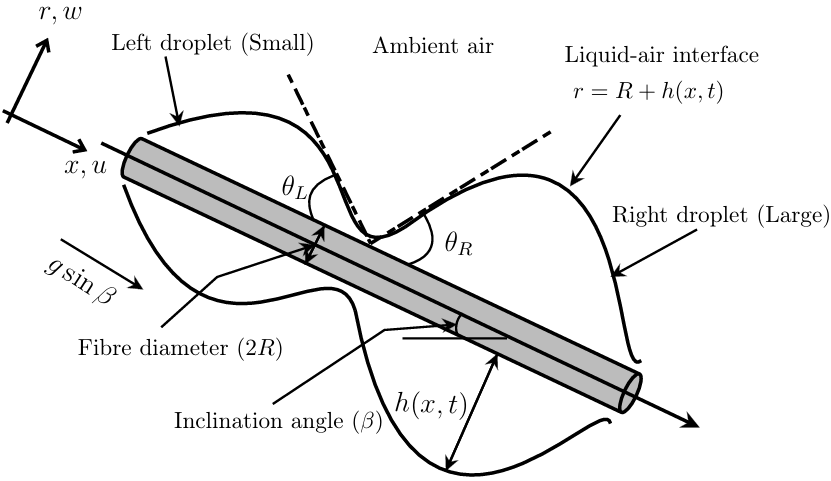}
    \caption{Schematic figure of two liquid droplets on a sloped cylinder. The left and right droplets are also referred to as the top and bottom droplets, respectively.}
    
    \label{fig:fiber_schematic}
\end{figure}
We consider a cylindrical fibre of radius $R^*$ inclined at an angle $\beta$ $\left(0\ll\beta\leq\pi/2\right)$ and the fibre is coated with an incompressible liquid film (see the configuration in Figure \ref{fig:fiber_schematic}). The density $\rho$, dynamic viscosity $\mu$, and surface tension $\sigma$ of the liquid are assumed to be constant. Following the work in \cite{craster2006viscous}, \cite{chan2021film} and \cite{zhao2023slip}, we treat the liquid-air interface as axisymmetric and neglect the effects of gravity in the normal direction. This assumption is valid when the inclination angle $\beta$ is large and the gravity in the normal direction is negligible compared to that in the streamwise direction.

\par The governing equations consist of the continuity equation and Navier-Stokes equations in cylindrical coordinates, where $x^*$ and $r^*$ represent the axial and radial coordinates, respectively. 
The velocity components in the $x^*$ and $r^*$ directions are denoted by $u^*$ and $w^*$, respectively. The axisymmetric liquid-air interface is described by $h^*(x^*,t^*)$, with the flow domain $R^* \leq r^* \leq R^*+h^*$. The dimensional equations of motion are:
\begin{subequations}\label{dim_model}
\begin{equation}\label{eq1}
u^*_{x^*}+\frac{1}{r^*}\left(r^*w^*\right)_{r^*}=0,
\end{equation}
\begin{equation}\label{eq2}
\rho\left(u^*_{t^*}+w^*u^*_{r^*}+u^*u^*_{x^*}\right)=-\left(p^*+\Pi^*\right)_{x^*}+\mu\left[\frac{1}{r^*}\left(r^*u^*_{r^*}\right)_{r^*}+u^*_{x^*x^*}\right]+\rho g\sin\beta,
\end{equation}
\begin{equation}\label{eq3}
\rho\left(w^*_{t^*}+w^*w^*_{r^*}+u^*w^*_{x^*}\right)=-p^*_{r^*}+\mu\left[\left(\frac{1}{r^*}\left(r^*w^*\right)_{r^*}\right)_{r^*}+w^*_{x^*x^*}\right],
\end{equation}
where $p^*$ is the liquid pressure, the term $\rho g \sin\beta$ represents the gravity in the streamwise direction, and $\Pi^*$ is the disjoining pressure describing intermolecular forces between the liquid film and the solid wall (\cite{bonn2009wetting}). 

\par Various forms of disjoining pressure have been used to model liquid dynamics on hydrophobic and hydrophilic substrates by augmenting the governing equations \citep{RevModPhys.57.827}. In this work, to model well-wetting liquids on superhydrophilic fibres, we adopt the dimensional disjoining pressure 
\begin{equation}
 \Pi^*(h^*)= -\frac{A'}{h^{*3}},
 \label{eq:disjoiningPressure}
\end{equation}
where $A'>0$ is a stabilization parameter. The functional form of $\Pi^*(h^*)$ is consistent with the long-range attractive component of the apolar van der Waals forces, which favors a thick stable liquid film. Similar formulations have been applied in studies of other droplet-fibre systems \citep{reisfeld1992non,ji2019dynamics,ji2021thermally,chattopadhyay2024modeling}.

\par At the surface of the fibre $r^*=R^*$, the no-slip and no-penetration boundary conditions are imposed as
\begin{equation}\label{eq4}
u^*=w^*=0.
\end{equation}
The boundary conditions at the liquid-air interface $r^*=R^*+h^*$ consist of the balance of normal and shear stresses, 
\begin{equation}\label{eq5}
\left(u^*_{r^*}+w^*_{x^*}\right)\left(1-h^{*2}_{x^*}\right)+2\left(w^*_{r^*}-u^*_{x^*}\right)h^*_{x^*}=0,
\end{equation}
\begin{equation}\label{eq6}
p^*_a-p^*+2\mu\left[w^*_{r^*}-\left(u^*_{r^*}+w^*_{x^*}\right)h^*_{x^*}+u^*_{x^*}h^{*2}_{x^*}\right]\left(1+h^{*2}_{x^*}\right)^{-1}=\sigma\kappa^*,
\end{equation}
and the kinematic condition
\begin{equation}\label{eq7}
w^*=h^*_{t^*}+u^*h^*_{x^*},
\end{equation}
\end{subequations}
where $\kappa^*=\left[h^*_{x^*x^*}-\frac{1}{r^*}\left(1+h^{*2}_{x^*}\right)\right]\left(1+h^{*2}_{x^*}\right)^{-3/2}$ is the local twice mean interfacial curvature and $p_a^*$ represents the atmospheric pressure.

\par To nondimensionalize the system (\ref{dim_model}), we introduce the dimensionless variables
\begin{equation}\label{eq8}
x^* = \mathcal Lx,~\left(r^*,h^*,R^*\right) = \mathcal H\left(r,h,R\right),~t^*=\mathcal Tt,~\left(u^{*}, w^{*}\right)=\mathcal U\left(u,\epsilon w\right),~
p^*=p^*_a+\mathcal Pp.
\end{equation}
By balancing the gravitational and viscous diffusion terms in \eqref{eq2}, we choose the characteristic streamwise velocity scale as $\mathcal U=g\mathcal H^{2}/\nu$, where $\nu=\mu/\rho$ is the kinematic viscosity.
The characteristic time scale is given by
$\mathcal T=\mathcal L/\mathcal U$, the pressure scale by $\mathcal P=\mu\mathcal U/\mathcal H$, and the characteristic length by $\mathcal L=\mathcal H/\epsilon$. The characteristic thickness $\mathcal{H}$ is taken to be the peak height of the larger droplet in the coalescence experiment, and the aspect ratio $\epsilon$ is assumed to satisfy $\epsilon  \ll 1$.  
Following the work of \cite{duprat2007absolute} and \cite{ruyer2008modelling},
we set the scale ratio $\epsilon = We^{-1/3}$, where $We = \sigma/\left(\mu\mathcal U\right) = \sigma/(\rho g\mathcal{H}^2)$ is the Weber number as conventionally defined in the falling-film literature \citep{kalliadasis2011falling}, representing the ratio of surface tension to the viscous normal stress generated by gravity at the free interface.  

In the limit $\epsilon \to 0$, applying (\ref{eq8}) to (\ref{dim_model}), we obtain the following leading-order dimensionless system
\begin{subequations}\label{dimless_model}
\begin{equation}\label{eq16}
u_x+r^{-1}\left(rw\right)_r=0,
\end{equation} 
\begin{equation}\label{eq17}
\epsilon Re\left(u_t+wu_r+uu_x\right)=\Omega-\epsilon p_x+\left(Sh^{-3}\right)_x+r^{-1}\left(ru_r\right)_r,
\end{equation}
\begin{equation}\label{eq18}
p_r = 0,
\end{equation}
\begin{equation}\label{eq19}
u=w=0\quad\text{at}~r=R,
\end{equation}
\begin{equation}\label{eq20}
u_r=0 \quad\text{at}~r=R+h,
\end{equation}
\begin{equation}\label{eq21}
-p=We\left(\epsilon^2h_{xx}-r^{-1}\right)\quad\text{at}~r=R+h,
\end{equation}
\begin{equation}\label{eq22}
w=h_t+uh_x\quad\text{at}~r=R+h,
\end{equation}
\end{subequations}
where $Re=\mathcal{UH}/\nu$ is the Reynolds number and the parameter $\Omega=\sin\beta$. Here, we have assumed that $\epsilon Re = O(1)$. Equation \eqref{eq18} implies that the pressure satisfies $p = p(x)$.
Moreover, since the aspect ratio $\epsilon = We^{-1/3} \ll 1$, the leading-order balance \eqref{eq21} can be rewritten as
\begin{equation}
    \epsilon p = \frac{1}{\eta (h+R)} - h_{xx},
\label{eq:pressure}
\end{equation}
where $\eta = \epsilon^2$. The right-hand side of \eqref{eq:pressure} represents a balance between the destabilizing azimuthal surface tension and the stabilizing streamwise surface tension of the free interface. 
The streamwise curvature term $h_{xx}$ is formally of higher order than the azimuthal curvature term $1/(\eta(h+R))$. However, as noted in previous studies \citep{craster2006viscous,ruyer2008modelling}, including the $h_{xx}$ term is essential for maintaining the correct regularization of the problem, as it contains the highest-order derivative and ensures consistency with long-wave theory.


The film stabilization term $Sh^{-3}$ in \eqref{eq17} originates from  the disjoining pressure \eqref{eq:disjoiningPressure}, where the constant $S=A'/\left(\mathcal{HL}\mu\mathcal U\right)$ is a dimensionless stabilization parameter. When describing the inertia-dominated dynamics in section \ref{sec:later_coal}, following \cite{ji2019dynamics}, we choose the stabilization parameter 
$S = ({\alpha \epsilon_p^3})/[{\eta(\alpha\epsilon_p + 1)]}>0$, which depends on the dimensionless thickness $\epsilon_p$ that characterizes the spatially-uniform steady state film thickness of the system. For the early-stage capillary-driven dynamics discussed in section \ref{sec:early_coal}, we set the stabilization parameter $S = 0$ to highlight the role of other physical mechanisms governing the early-stage evolution.


\par We consider the Nusselt uniform film without perturbations of the free surface, for which the base flow velocity $\widehat u(r)$ is determined by the balance between gravity and wall friction, satisfying the equation
\begin{subequations}\label{eq24}
\begin{equation}
L_d\left(\widehat u\right)=-\Omega, \quad \mbox{where }L_d(\widehat u):=r^{-1}\left(r\widehat u_r\right)_r,
\end{equation}
with boundary conditions 
\begin{equation}
\widehat u=0 \mbox{ at } r=R, \qquad \widehat u_r=0 \mbox{ at } r=R+h. 
\label{eq:nusselt_BC}
\end{equation}
\end{subequations}
Here, $L_d(\widehat u)$ denotes a linear differential operator that accounts for wall friction.
Solving the boundary value problem (\ref{eq24}) leads to Nusselt flow velocity profile
\begin{equation}\label{eq25}
\widehat u=\Omega\left[\frac{(R+h)^2}{2}\ln\left(\dfrac{r}{R}\right)-\frac{1}{4}\left(r^2-R^2\right)\right].
\end{equation}

Next, we follow \cite{ruyer2008modelling} to derive the weighted-residual model (WRM) for the evolution of the film thickness $h(x,t)$. The key idea of this approach is to project the streamwise velocity field $u(x,t)$ onto an appropriately chosen weight function and then apply the Galerkin weighted residual method.

We consider an asymptotic expansion of the velocity $u$ of the form $u=\sum_{j=0}^{N}\epsilon^{j}u_{j}$, where the leading-order term is given by $u_0 = \mathcal{A}_0\vartheta_{0}$ with $\mathcal{A}_0 = O(1)$. The weight function, chosen as $\mathcal{W}=\vartheta_0$, is a polynomial in $r$ of the same degree as the Nusselt velocity $\widehat u$ in (\ref{eq25}). Moreover, this weight function must satisfy the boundary conditions outlined in \eqref{eq:nusselt_BC}, namely $\mathcal{W}\vert_{r=R}=\mathcal{W}_{r}\vert_{r=R+h}=0$. Consequently, we select $\mathcal{W}$ to adopt the same functional form as $\widehat u$ in (\ref{eq25}), and the leading-order approximation of the velocity becomes $u_0 = \mathcal{A}_0\widehat u$.

Substituting the expression for the pressure field \eqref{eq:pressure} into
\eqref{eq17} yields the leading-order streamwise momentum equation
\begin{equation}\label{eq23}
\delta\left(u_{t}+wu_r+uu_x\right)=\mathbb A[h]+r^{-1}\left(ru_r\right)_r,
\end{equation}
where $\delta = \epsilon Re$ is the reduced Reynolds number and the operator $\mathbb A$ is defined as
\begin{equation}
    \mathbb A[h]=\Omega+h_{xxx}+\frac{h_x}{\eta(R+h)^2}+\left(\frac{S}{h^3}\right)_x.
\end{equation}
To apply the Galerkin weighted residual method to \eqref{eq23}, we use the weight function $\mathcal{W}(r)$ together with the leading-order approximation $u = u_0$. The residual of the approximation is required to be orthogonal to the weight function, yielding
\begin{equation}\label{eq37}
\delta\int_R^{R+h}\left(u_{0t}+w_0u_{0r}+u_0u_{0x}\right)r\mathcal{W}(r)\text{d}r=\mathbb A[h]\int_R^{R+h}r\mathcal{W}(r)\text{d}r+\int_R^{R+h}L_d(u_0)r\mathcal{W}(r)~\text{d}r.
\end{equation}

\par The kinematic condition (\ref{eq22}), combined with the continuity equation (\ref{eq16}) and the boundary conditions (\ref{eq19}), can be expressed as a mass conservation equation 
\begin{equation}\label{eq29}
(1+\alpha h)h_t+q_x=0, \qquad q=\frac{1}{R}\int_R^{R+h}ur~\text{d}r,
\end{equation}
where $q$ represents the local flow rate and $\alpha=1/R$ is the aspect ratio between the dimensional characteristic film thickness $\mathcal{H}$ and the dimensional fibre radius $R^*$.

Using (\ref{eq25}) together with the definition of $q$ in \eqref{eq29}, and approximating the velocity as $u \approx u_0 = \mathcal{A}_0\widehat{u}$, we obtain the expression for $\mathcal A_0$,
\begin{equation}\label{eq30}
\mathcal A_0=\frac{3q}{h^3\Omega\varphi(\alpha h)},
\end{equation}
where the shape factor $\varphi$ is defined by
\begin{equation}\label{eq31}
\varphi(\zeta)=\frac{3}{16\zeta^3}\left[\left(1+\zeta\right)^4\left\{4\text{ln}(1+\zeta)-3\right\}+4\left(1+\zeta\right)^2-1\right].
\end{equation}
Using (\ref{eq30}), the leading-order velocity profile $u_{0}$ can be written as
\begin{equation}\label{eq32}
u_{0}=\frac{3q}{h\varphi(\alpha h)}\left[\left(\frac{1}{2}+\frac{1}{\alpha h}+\frac{1}{2\left(\alpha h\right)^2}\right)\ln(\alpha r)-\frac{r-R}{2h^2}\left(\frac{r-R}{2}+\frac{1}{\alpha}\right)\right].
\end{equation}

Substituting \eqref{eq32} and $\mathcal{W} = \widehat{u}$ into \eqref{eq37}
and combining the result with the mass conservation equation and the definition of $q$ in \eqref{eq29}, we obtain the first-order weighted-residual model
\begin{subequations}
\label{eq:WRM_main}
\begin{equation}\label{eq:continuity}
(1+\alpha h)h_t+q_x=0,
\end{equation}
\begin{equation}
\label{eq:q_equation}
\delta\left(q_t+\Theta_1(\alpha h)\frac{qq_x}{h}-\Theta_2(\alpha h)\frac{q^2h_x}{h^2}\right)=I(\alpha h)\left[h\left\{\Omega-\left(\mathcal Z(h)-h_{xx}\right)_x\right\}-\frac{3 q}{h^2 \varphi(\alpha h)}\right],
\end{equation}
where the auxiliary functions $\mathcal{Z}, \Theta_1, \Theta_2, M, N$, and $I$ are defined as
\begin{equation}
    \mathcal Z(h)=\frac{\alpha}{\eta(1+\alpha h)}-\frac{S}{h^3}, \quad \Theta_1(\zeta)=\frac{3\left[(1+\zeta)^2\chi(\zeta)-10\right]}{M(\zeta)}, \quad
    \Theta_2(\zeta)=\frac{\Upsilon(\zeta)}{4N(\zeta)},
\end{equation}
\begin{equation}
    M(\zeta)=16\zeta^2\varphi(\zeta)\psi(\zeta), \quad N(\zeta)=\zeta^2M(\zeta)\varphi(\zeta), \quad I(\zeta)=\frac{64\zeta^5\varphi^2(\zeta)}{3\psi(\zeta)}.
\end{equation}
\end{subequations}
The functions $\chi(\zeta)$, $\psi(\zeta)$, $\Upsilon(\zeta)$ are defined in Appendix \ref{appendix:a}. 
The derivation of \eqref{eq:q_equation} is lengthy but straightforward. A closely related derivation can be found in \cite{ruyer2008modelling} and \cite{novbari2018parametric}.

\par The system \eqref{eq:WRM_main} models droplet dynamics on a sloped cylindrical surface, incorporating the effects of moderate inertia, surface tension, gravity, and intermolecular forces. In equation \eqref{eq:q_equation}, the influence of inertia is represented by the left-hand-side terms associated with the reduced Reynolds number $\delta$. On the right-hand side, $\mathcal Z(h)$ captures the destabilizing effect of azimuthal curvature, while $h_{xx}$ corresponds to the stabilizing influence of streamwise curvature. Additionally, $\mathcal Z(h)$ includes a film stabilization term $S/h^3$ which ensures complete wetting of the cylinder. The term $\Omega$ characterizes the dynamics driven by gravity through its streamwise component,  and the term $3q/[h^2\varphi(\alpha h)]$ describes the viscous drag within the film.
For a vertical cylinder $\left(\beta=\pi/2\right)$,  we have $\Omega=1$, and the model \eqref{eq:WRM_main} reduces to the first-order weighted-residual model derived by \cite{ruyer2008modelling} when the stabilization term is neglected $(S = 0)$. Although \cite{ruyer2008modelling} also developed a second-order weighted-residual model, in this work we restrict our attention to the first-order formulation given by \eqref{eq:WRM_main}.

\par The intricate dynamics of liquid droplets on fibres have also been explored using other simplified, idealized models in which the droplets are treated as solids or point masses. For example, \cite{poulain2023sliding} used a forced elastic pendulum model to analogously describe the vibration and swing of droplets on an oscillating fibre. The recent work of  \cite{feng2024short} introduced a mass-spring-damper (MSD) model to explain the dynamics of two coalescing droplets on a cylindrical fibre. In the MSD model, the damping term represents viscous friction that resists the relative motion of the droplets along the fibre. 
In contrast, within our continuum formulation \eqref{eq:WRM_main} for the liquid-air surface, the frictional effects arise naturally from the viscous dissipation term $r^{-1}(ru_r)_r$ in the streamwise momentum equation, which governs the velocity profile of the Nusselt base flow \citep{ruyer2008modelling}. 

Next, in section \ref{sec:early_coal}, we discuss a simplified lubrication model obtained as a limiting case of the system \eqref{eq:WRM_main}.
The weighted-residual model \eqref{eq:WRM_main} for droplet dynamics with moderate inertia effects will be studied in section \ref{sec:later_coal}.

\section{Early-stage coalescence: capillary-driven dynamics}
\label{sec:early_coal} 
At short times immediately after the droplets meet, the radius of the liquid bridge connecting the two droplets is close to the fibre radius $R^*$.
In this regime, the bridge is sufficiently thin and its dynamics are dominated by highly localized surface tension, while inertia and gravity play negligible roles.
\cite{hernandez2012symmetric} showed that for asymmetric coalescence on a flat substrate, the shape of the liquid bridge can be described by a similarity solution of a one-dimensional lubrication equation, and the asymmetry in the left and right adjoining contact angles \citep{karpitschka2014coalescence,Karpitschka_Riegler_2014} can induce a lateral displacement of the bridge minimum toward the side with the smaller adjoining contact angle. In this section, we show that the early-stage coalescence of two droplets of unequal sizes on a cylindrical fibre asymptotically exhibits similar behaviors.

\par We consider a meniscus bridge profile $h(x,t)$ connecting two droplets with its left and right adjoining contact angles, $\theta_L$ and $\theta_R$ (see Figure~\ref{fig:fiber_schematic}), respectively.
Assuming that the inertia, gravity, and film stabilization terms are neglected, $\delta\to 0$, $\Omega \to 0$, and $S\to 0$,
we reduce the coupled first-order WRM equations \eqref{eq:WRM_main} to the lubrication equation for the meniscus profile $h(x,t)$,
\begin{equation}
\left(h+\frac{\alpha}{2}h^2\right)_t+\left[\mathcal M(h)\left(-\frac{\alpha}{\eta(1+\alpha h)}+h_{xx}\right)_x\right]_x=0,
\label{eq:lubrication}
\end{equation}
where the mobility function $\mathcal{M}(h)={h^3}\varphi(\alpha h)/3$. 
In the limit $h \to 0$, we have
\begin{equation}
    \mathcal{M}(h) = \frac{h^3}{3}(1+\alpha h) + O(h^5).
\label{eq:mobility_leading}
\end{equation}
Similar lubrication models have been widely studied in prior work for droplet-fibre systems in the inertialess limit \citep{craster2006viscous,haefner2015influence,zhao2024inertia}. The dimensionless lubrication model \eqref{eq:lubrication} can also be derived directly from classical lubrication theory \citep{oron1997long}. In this formulation, we set the aspect ratio $\epsilon = \mathcal{H}/\mathcal{L} = \theta_L$, which implies $\eta = \epsilon^2 = \theta_L^2$. The characteristic axial velocity, time, and pressure scales are chosen as $\mathcal{U} = ({\sigma}\epsilon^3)/{\mu}$,  $\mathcal{T} = \mathcal{L}/\mathcal{U}$, and  $\mathcal{P} = (\mu \mathcal{U})/(\epsilon\mathcal{H})$, respectively.

\par Following the approach in \cite{hernandez2012symmetric} for asymmetric coalescence of two droplets on a flat substrate, 
we seek asymptotic similarity solutions of \eqref{eq:lubrication} 
using the ansatz
\begin{equation}
 1+\alpha h(x,t) = (1+\alpha h_0(t)) \F(\xi), \quad    \xi = \frac{\alpha(x-x_0(t))}{1+\alpha h_0(t)}, 
 \label{eq:ansatz}
\end{equation}
where $\F(\xi)$ is the leading-order similarity profile of the meniscus bridge, $x_0(t) = \arg\min_{x} h(x,t)$ represents the horizontal location of bridge minimum during coalescence, and $h_0(t)$ represents the scale of the elevated bridge height at its minimum in time.

Substituting the ansatz \eqref{eq:ansatz} into \eqref{eq:lubrication} and retaining the leading-order terms in the mobility function \eqref{eq:mobility_leading}, we obtain the leading-order equation,
\begin{equation}
\F\left[h_0'\F-\F_{\xi}\left(\xi h_0'+x_0'\right)\right]+\left[\frac{\F\left[(1+\alpha h_0)\F-1\right]^3}{3(1+\alpha h_0)^3}\left(\frac{\F_{\xi}}{\theta_L^2\F^2}+\F_{\xi\xi\xi}\right)\right]_{\xi} = 0,
\label{eq:sim_full}
\end{equation}
where the first term originates from the time derivative, the remaining terms represent the contributions from azimuthal and streamwise curvatures, and primes $(')$ denote differentiation with respect to time.

Setting
\begin{equation}
    h_0' = \frac{1}{3(1+\alpha h_0)^3}, \quad x_0' = \frac{U}{3(1+\alpha h_0)^3}={Uh_0^{\prime}},
\label{eq:migration_ode}
\end{equation}
where $U$ is an unknown constant to be determined, and assuming that $\alpha h_0 \ll 1$ during the early stage of coalescence,  we reduce \eqref{eq:sim_full} to the leading-order similarity equation for $\F(\xi)$,
\begin{subequations}
\begin{equation}
\F\left[\F-\left(\xi +U\right)\F_{\xi}\right]+\left[{\F(\F-1)^3}\left(\frac{\F_{\xi}}{\theta_L^2\F^2}+\F_{\xi\xi\xi}\right)\right]_{\xi} = 0.
\label{eq:similarity}
\end{equation}
The boundary conditions imposed by the prescribed adjoining contact angles at the left and right edges of the bridge, 
$h^*_{x^*}=\theta_L$ and $h^*_{x^*}=\theta_R$, lead to asymptotic far field boundary conditions for $\F$,
\begin{equation}
\F_{\xi}(-\infty) = -1, \quad \F_{\xi}(\infty) = \frac{\theta_R}{\theta_L}.
\label{eq:sim_bc}
\end{equation}
Moreover, at the location of the bridge minimum, where $h_{x}(x_0(t),t)=0$, we obtain
\begin{equation}
\F_{\xi}(0) = 0.
\label{eq:sim_minCondition}
\end{equation}
\label{eq:sim}
\end{subequations}
\par We note that, unlike coalescence on a flat substrate where the lubrication equation admits an exact self-similar solution \citep{hernandez2012symmetric}, the cylindrical geometry introduces an azimuthal curvature term associated with the intrinsic length scale of the fibre radius, thereby breaking the exact scaling invariance of the governing equation. Consequently, the present asymptotic analysis identifies the corresponding asymptotic self-similar solution and shows that the fibre curvature contributes to the leading-order similarity equation \eqref{eq:sim}.

In the limit $\alpha \to 0$, the ODEs \eqref{eq:migration_ode} reduce to $h_0'=1/3$ and $x_0' = U/3$, indicating constant migration speeds of the bridge minimum in both the vertical and horizontal directions. This behavior is consistent with the results reported in \cite{hernandez2012symmetric} for asymmetric droplet coalescence on a flat substrate. In contrast, with a finite positive $\alpha$, the fibre geometry influences the dynamics of the liquid bridge, leading to nonlinear behavior in the growth and migration of the bridge.

\begin{figure}
\centering
\subfloat[]{    \includegraphics[height=4.5cm]{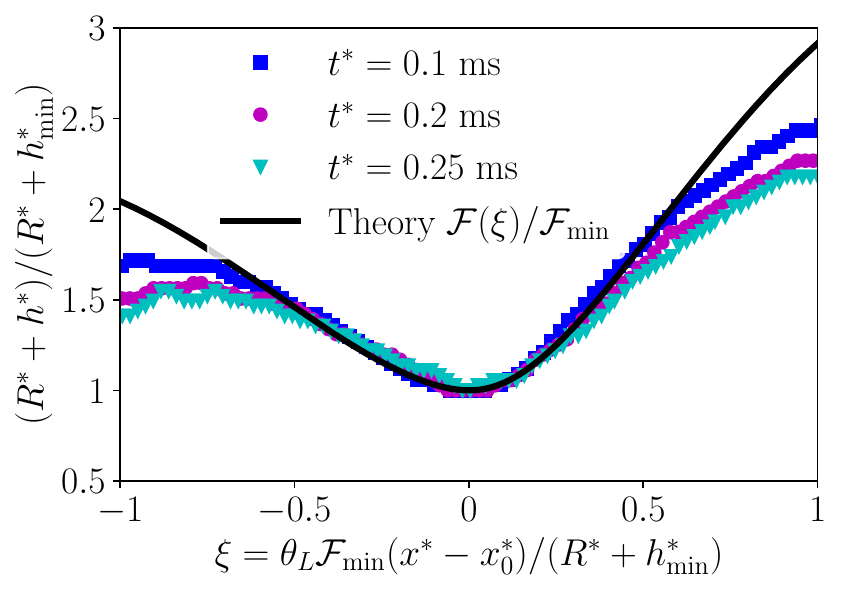}}
\subfloat[]{  \includegraphics[height=4.5cm]{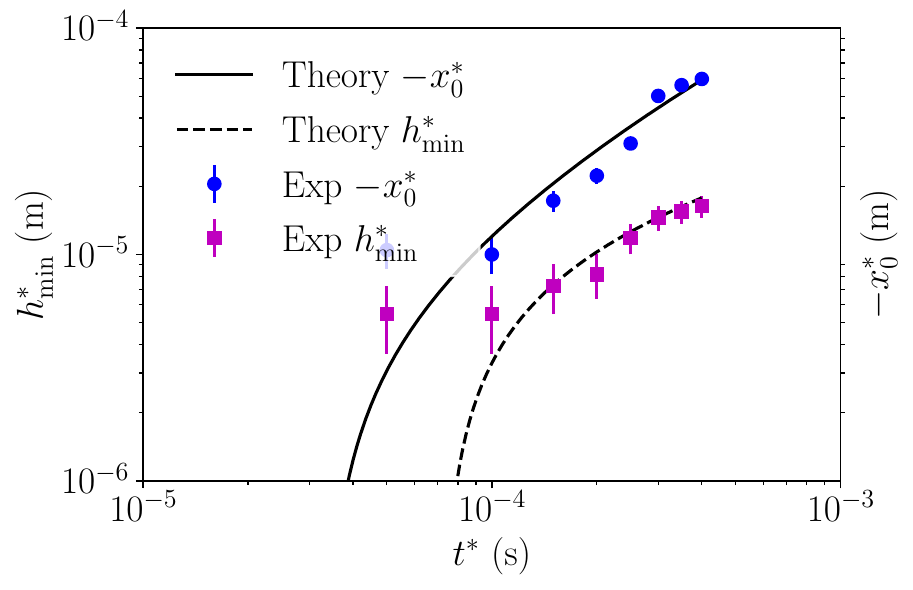}}
    \caption{(a) Rescaled experimental profiles of an asymmetric droplet coalescence event at short times after initial connection, compared with the similarity profile obtained by numerically solving \eqref{eq:sim}; (b) Horizontal and vertical positions of the meniscus bridge during early-stage coalescence, compared with theoretical predictions in \eqref{eq:migration_pred}. }
    \label{fig:sim_coalesence}
\end{figure}
To quantitatively assess the asymptotic similarity solution, we compare experimental observation against the analytical predictions from the similarity ODE boundary value problem \eqref{eq:sim}.
Figure \ref{fig:sim_coalesence} presents rescaled experimental profiles and the corresponding locations of the meniscus bridge from the axisymmetric droplet coalescence experiment shown in Figure \ref{fig:experiment_90}, where the two droplets are deposited on a vertical fibre. The smaller droplet on the left (top in Figure \ref{fig:experiment_90}) has an adjoining contact angle of $\theta_L = 13.2^{\circ}$, and the larger droplet on the right (bottom in Figure \ref{fig:experiment_90}) has an adjoining contact angle of $\theta_R = 26.7^{\circ}$. 

In Figure \ref{fig:sim_coalesence} (a), the bridge profiles ${R}^*+h^*$ from experimental observations at different times are rescaled by ${R}^*+h^*_{\min}$ and plotted as functions of the similarity variable $\xi = \theta_L\F_{\min}(x^*-x^*_0(t^*))/({R}^*+h_{\min}^*(t^*))$ (see solid markers). Here, $x^*_0(t^*)$ and $h^*_{\min}(t^*)$ represent the dimensional position and thickness of the lowest point of the meniscus bridge in physical time $t^*$, $\F_{\min}$ is the minimum of the similarity solution $\F(\xi)$, and the formula of $\xi$ is consistent with the definition in \eqref{eq:ansatz} using $\alpha = \mathcal{H}/R^*$, and $\epsilon = \theta_L = \mathcal{H}/\mathcal{L}$. 

The black solid curve in Figure \ref{fig:sim_coalesence} (a) represents the similarity solution $\F(\xi)$ obtained by numerically solving the similarity equation \eqref{eq:sim} over the domain $-1\le \xi \le 1$. Specifically, we use finite differences and Newton's iteration to solve for both the similarity profile $\mathcal{F}(\xi)$ and the constant $U$. 
The numerical results suggest that the boundary value problem \eqref{eq:sim} has a unique similarity solution $\F(\xi)$ associated with $U = -3.18$, and the minimum value of $\F(\xi)$ is $\F(0) = \F_{\min}=1.17$. This is similar to the results for droplet coalescence on a flat surface studied in \cite{hernandez2012symmetric}.
The rescaled experimental profiles align well with the analytical self-similar solution, indicating that the similarity equation \eqref{eq:sim} accurately captures the dynamics of the meniscus bridge during early-stage coalescence. 

We further observe that the growth and horizontal migration of the bridge minimum are characterized by the coupled ODEs \eqref{eq:migration_ode}. Taking the spatial minimum of the first equation in \eqref{eq:ansatz}, we obtain the relation $h_{\min} = \tfrac{1}{\alpha}[(1+\alpha h_0)\F_{\min}-1]$.
Using this expression, the system \eqref{eq:migration_ode} can be solved explicitly, yielding
\begin{equation}
    h_{\min}(t) = \F_{\min}\left(\frac{C+4t}{3\alpha^3}\right)^{1/4}-\frac{1}{\alpha},
     \quad x_0(t) = U\left(\frac{C+4t}{3\alpha^{3}}\right)^{1/4}+D,
\label{eq:migration_pred}
\end{equation}
where $C$ and $D$ are constants of integration determined by the initial conditions $x_0=0$ and $h_{\min}=0$ at time $t = 0$ when the two droplets first come into contact.

Figure \ref{fig:sim_coalesence} (b) shows the temporal evolution of the dimensional horizontal displacement ($-x^*_0$) and the thickness ($h^*_{\min}$) of the bridge (blue and purple markers) on a log-log scale. The results indicate that the bridge migrates toward the smaller droplet, corresponding to $x^*_0 < 0$, as the bridge thickness $h^*_{\min}$ increases with time. The experimental data show good agreement with the analytical predictions (solid and dashed curves) given by \eqref{eq:migration_pred}, which supports the conclusions of our lubrication-based model for describing early-stage asymmetric coalescence between droplets of unequal sizes.

Before turning to the analysis of late-stage coalescence, we briefly discuss the case of coalescence between two equally sized droplets. In this situation, where the adjoining contact angles $\theta_L = \theta_R = \theta$, we expect the coalescence process to be symmetric and no horizontal migration occurs. That is, the bridge grows without lateral displacement, corresponding to $U = 0$ in the similarity formulation.
Under these conditions, the boundary value problem for the leading-order similarity equation \eqref{eq:sim} reduces to
\begin{subequations}
\begin{equation}
\F\left(\F-\xi\F_{\xi}\right)+\left[{\F(\F-1)^3}\left(\frac{\F_{\xi}}{\theta^2\F^2}+\F_{\xi\xi\xi}\right)\right]_{\xi} = 0, 
\label{eq:similarity_symmetric}
\end{equation}
\begin{equation}
\F_{\xi}(\infty) = 1, \quad \F_{\xi}(0) = 0, \quad \F_{\xi\xi\xi}(0) = 0,
\label{eq:sim_bc_symmetric}
\end{equation}
\label{eq:sim_symmetric}
\end{subequations}
where the additional condition $\F_{\xi\xi\xi}(0) = 0$ follows from the symmetry of the similarity solution about the center $\xi = 0$. 
\begin{figure}
\centering
\subfloat[]{    \includegraphics[height=4.5cm]{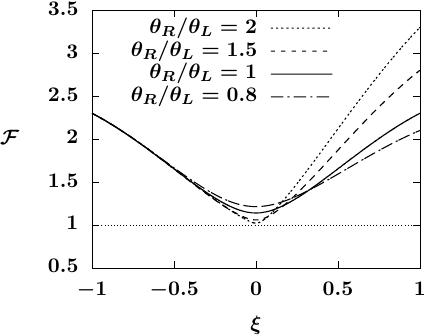}}\qquad
\subfloat[]{  \includegraphics[height=4.5cm]{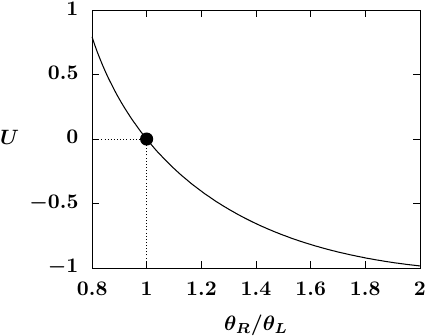}}
\caption{(a) Typical similarity solutions $\mathcal{F}(\xi)$ of the ODE \eqref{eq:sim} for $\theta_L = \pi/9$ and varying adjoining contact angle ratios $\theta_R/\theta_L$. (b) The corresponding values of $U$ as a function of the ratio $\theta_R/\theta_L$ showing that $U=0$ for $\theta_R = \theta_L$.
}
\label{fig:sim_continuation}
\end{figure}

Figure \ref{fig:sim_continuation} (a) presents a family of similarity solutions $\mathcal{F}(\xi)$ obtained by solving the similarity ODE \eqref{eq:sim} for asymmetric coalescence with a fixed left adjoining contact angle $\theta_L = \pi/9$ and varying right adjoining contact angles $\theta_R$. A parametric continuation method is used to numerically trace this family of solutions.
For $\theta_R/\theta_L = 1$, the similarity profile $\mathcal{F}(\xi)$ (solid curve) becomes symmetric about $\xi = 0$ and satisfies the ODE problem \eqref{eq:sim_symmetric} for symmetric coalescence. 
For $\theta_R/\theta_L \neq 1$, the effect of asymmetric adjoining contact angles is incorporated through the boundary condition \eqref{eq:sim_bc}, which governs the behavior of the similarity profile $\mathcal{F}$ as $\xi \to \infty$.
Both systems \eqref{eq:sim} and \eqref{eq:sim_symmetric} become degenerate for $\mathcal{F} = 0$ and $\mathcal{F} = 1$. However, the numerical solutions shown in Figure \ref{fig:sim_continuation} (a) satisfy $\mathcal{F}(0) = \mathcal{F}_{\min} > 1$, indicating that they are physically relevant similarity profiles corresponding to nonnegative bridge thicknesses. 

Figure~\ref{fig:sim_continuation} (b) shows the dependence of the migration velocity parameter $U$ on the adjoining contact angle ratio $\theta_R/\theta_L$. The results demonstrate that $U$ decreases monotonically as the ratio increases. Specifically, when $\theta_R < \theta_L$, we find $U > 0$, indicating that the liquid bridge migrates toward the smaller droplet on the right. In contrast, when $\theta_R > \theta_L$, we have $U < 0$, and the liquid bridge migrates toward the smaller droplet on the left. For the symmetric case $\theta_R = \theta_L = \theta$, we obtain $U = 0$, which corresponds to no directional migration. These results confirm that the bridge consistently migrates toward the smaller droplet during the early-stage coalescence and that the migration speed strongly depends on the asymmetry in the adjoining contact angles.

\section{Late-stage coalescence: inertia-influenced dynamics}
\label{sec:later_coal}
In this section, we focus on the late-stage coalescence dynamics governed by the weighted-residual boundary layer (WRM) model \eqref{eq:WRM_main}. We first discuss the shape of nontrivial hydrostatic droplets on a cylindrical fibre in subsection \ref{sec:shape}, which will be used as the initial profile for numerical studies in subsection \ref{sec:numerics}. Subsection \ref{sec:migration_hydrodynamic} provides a hydrodynamic interpretation of the droplet migration mechanism based on the WRM model. The effects of varying the parent droplet height ratios and the inclination angle of the fibre will be discussed in subsection \ref{sec:ratio_inclination}.

\subsection{Quasi-static droplet profiles}
\label{sec:shape}
While droplets on a flat superhydrophilic substrate spread indefinitely to form a uniform film, those on a superhydrophilic cylinder can form a stable nontrivial hydrostatic shape due to the curvature of the cylindrical surface \citep{carroll1976accurate}.
Quasi-static droplets on a fibre experiencing negligible inertial and gravitational effects 
are described by the nontrivial steady states, $\bar{h}(x)$, of the WRM model \eqref{eq:WRM_main} with $\delta=\Omega=0$, satisfying
\begin{equation}
    \bar{h}_{xx} = \mathcal{Z}\left(\bar{h}\right)-\bar{P}, \quad \mbox{where } \mathcal{Z}\left(\bar{h}\right) = \frac{\alpha}{\eta\left(1+\alpha \bar{h}\right)} - \frac{S}{\bar{h}^3}.
\label{eq:SS}
\end{equation}
The second-order ODE \eqref{eq:SS} describes the balance of azimuthal and streamwise curvatures, and the stabilization term, where $\bar{P} > 0$ represents a spatially constant pressure. 

\begin{figure}
\centering
\subfloat[]{\includegraphics[height=4.5cm]{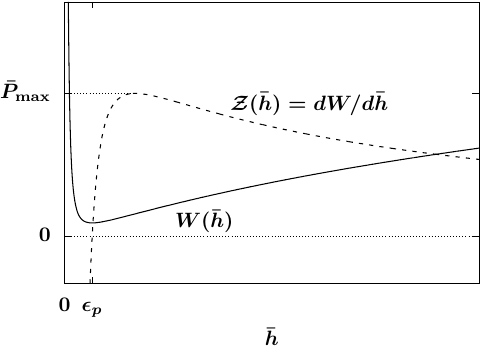}}
\qquad
\subfloat[]{ \includegraphics[height=4.5cm]{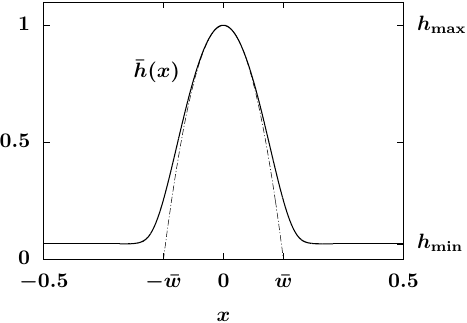}}
    \caption{(a) Plot of the potential $W\left(\bar{h}\right)$ and the combined pressure $\mathcal{Z}\left(\bar{h}\right)$; (b) A hydrostatic droplet solution $\bar{h}(x)$ corresponding to $\bar{P}<\bar{P}_{\max}$. The dashed curve represents the parabolic approximation \eqref{eq:parabola} for the core region of the droplet.}
    \label{fig:singleDrop}
\end{figure}

It is useful to define the potential 
\begin{equation}
W\left(\bar{h}\right) = \int \mathcal{Z}\left(\bar{h}\right)~\text{d}\bar{h} = \frac{S}{2\bar{h}^2}+\frac{1}{\eta}\ln\left(1+\alpha \bar{h}\right).
\end{equation}
With the stabilization parameter $S$ taking the value $S = ({\alpha \epsilon_p^3})/[{\eta(\alpha\epsilon_p + 1)]}$, where $\epsilon_p$ represents the precursor film thickness, we have $\mathcal{Z}\left(\epsilon_p\right)=0$. 
Consequently, the potential $W\left(\bar{h}\right)$ has a unique minimum at $\bar{h}=\epsilon_p$ and is similar to the potential for dewetting films \citep{glasner2003coarsening}. 
The function $\mathcal{Z}\left(\bar{h}\right)$ has a unique maximum at $\bar{h}_{\text{peak}}$, where $\mathcal{Z}\left(\bar{h}_{\text{peak}}\right) =\bar{P}_{\max}$ (see Figure~\ref{fig:singleDrop} (a)).
For $0 < \bar{P} <  \bar{P}_{\max}$, the ODE \eqref{eq:SS} admits a spatially uniform steady state $\bar{h} \equiv h_{\min}\left(\bar{P}\right) < \bar{h}_{\text{peak}}$, which satisfies $\mathcal{Z}\left(h_{\min}\right)=\bar{P}$.
In the limit of $\bar{P}\to 0$, we have
\begin{equation}
    h_{\min} = \epsilon_p + \frac{\epsilon_p\eta (\alpha \epsilon_p+1)^2}{\alpha(2\alpha \epsilon_p+3)}\bar{P}+O\left(\bar{P}^2\right).
\end{equation}

\par 
A continuous family of nontrivial isolated hydrostatic droplet profiles, denoted by $\bar{h}(x;\bar{P})$, exist as solutions to the ODE \eqref{eq:SS}, parameterized by the constant pressure $\bar{P}$ for $0 < \bar{P} < \bar{P}_{\max}$. 
For any fixed finite pressure $\bar{P}$ within the range, and with the minimum film thickness of the droplet set by $h_{\min}$, 
the first integral of equation \eqref{eq:SS} is given by
\begin{equation}
    \tfrac{1}{2}\left(\bar{h}_x\right)^2 =  W\left(\bar{h}\right)-W\left(h_{\min}\right)-\bar{P}\left(\bar{h}-h_{\min}\right).
\label{eq:firstIntegral}
\end{equation}
At the maximum of the droplet, we have $\bar{h}_x = 0$, which combined with \eqref{eq:firstIntegral}, leads to the condition for determining the height of the droplet $h_{\max}$, 
\begin{equation}
W\left(h_{\max}\right)-\bar{P}h_{\max} = W(h_{\min})-\bar{P}h_{\min}.
\label{eq:Rmax=0}
\end{equation}

For simplicity, we consider a nontrivial droplet centered at $x = 0$. The core region of the droplet takes a parabolic shape and can be approximated by
\begin{equation}
    \bar{h}(x;\bar{P}) \approx A\left(\bar{w}^2-x^2\right),
\label{eq:parabola}
\end{equation}
where the parameter $\bar{w}$ is the half-width of the droplet. Using the approximation \eqref{eq:parabola} at $x = 0$, we obtain the maximum of the droplet $h_{\max} = A\bar{w}^2$ which leads to $\bar{w} = \sqrt{h_{\max}/A}$. Matching the linearized curvature of the approximate profile \eqref{eq:parabola} and $\bar{h}_{xx}$ at $h = h_{\max}$ from the ODE \eqref{eq:SS} yields $A = \left(\bar{P}-\mathcal{Z}\left(h_{\max}\right)\right)/2$.

Figure~\ref{fig:singleDrop} (b) presents a typical nontrivial droplet solution $\bar{h}(x)$ over the finite domain $-0.5\le x \le 0.5$ obtained by numerically solving the ODE~\eqref{eq:SS} with aspect ratio parameters $\alpha=2$, $\eta=0.0049$ and pressure $\bar{P} = 208.7$. For all numerical results presented in this section, 
the stabilization parameter is set to $S = 0.047$, which corresponds to the precursor layer thickness $\epsilon_p = 0.05$. The resulting solution has a peak height $h_{\max} = 1$ and a minimum value $h_{\min}=0.067\gtrsim \epsilon_p$
 The figure also includes the corresponding parabolic approximation \eqref{eq:parabola}, demonstrating that the parabola accurately captures the droplet profile in the core region for $-\bar{w} < x < \bar{w}$.

\subsection{Droplet migration driven by coalescence with inertia effects}
\label{sec:numerics}
\par
\begin{figure}
    \centering
     \includegraphics[width=\linewidth]{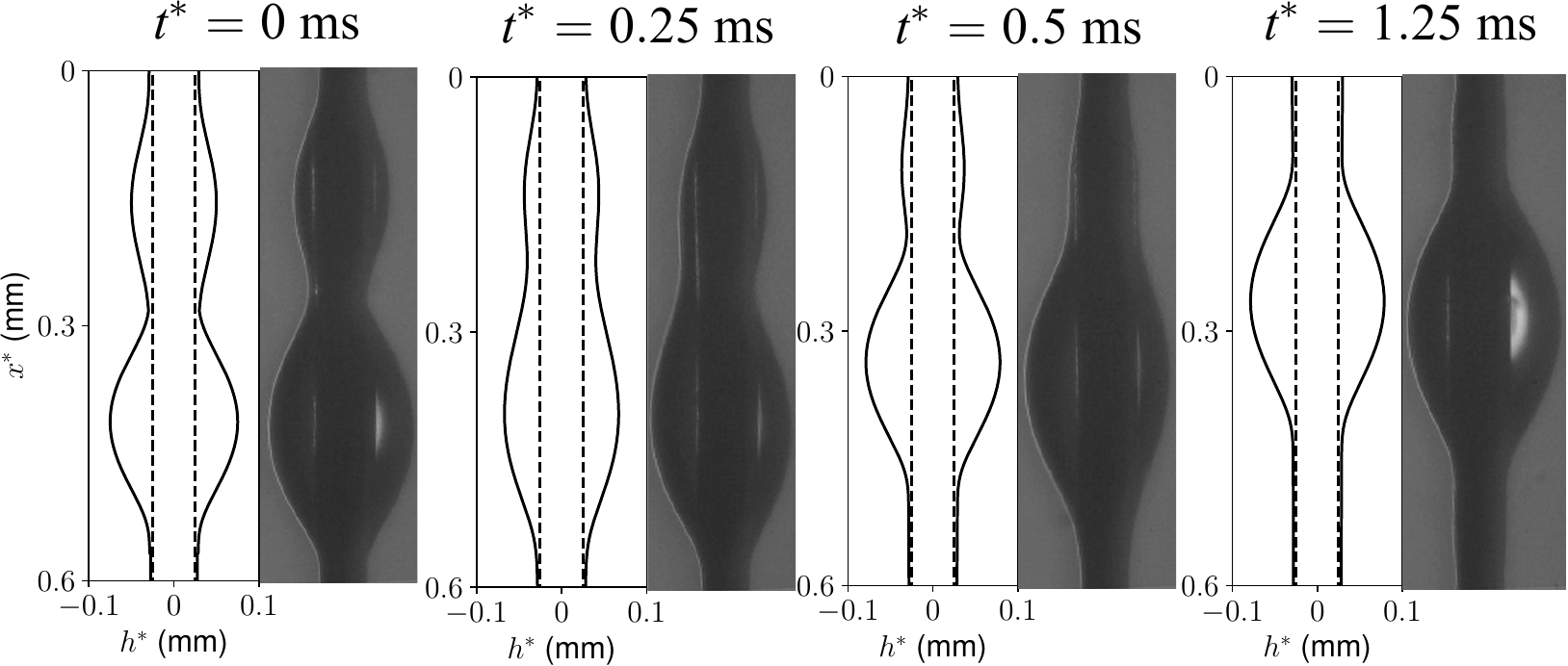}
    \caption{Comparison between the simulation of the WRM model \eqref{eq:WRM_main} (left panels) and experimental observations (right panels) for the coalescence of two droplets on a vertical fibre ($\beta=\pi/2$) at different times. The initial droplet configuration at $t^*=0$ ms is given by \eqref{eq:ic}. The other system parameters are $\alpha=2$, $\delta=0.085$, $\Omega=1$, $\eta=0.0049$, $ S=0.047$. In the numerical results, the solid curves denote the free surface, and the dashed lines indicate the fibre walls. A corresponding animation is provided in the supplementary movie (movie1.mp4).}
    \label{fig:coalesence_angle=90}
\end{figure}

 For the PDE simulations in the rest of the section, we set the initial droplet configuration as follows:
\begin{equation}
    h(x,0) = \max\left\{ \bar{h}\left(x-X_T;\bar{P}_T\right),\ \bar{h}\left(x-X_B;\bar{P}_B\right),\ \epsilon_p \right\}, ~q(x,0) = \frac{\epsilon_p^3\Omega}{3}\varphi(\alpha \epsilon_p), ~ 0 \le x \le L,
\label{eq:ic}
\end{equation}
where $X_T$ and $X_B$ denote the locations of the peaks of the top and bottom droplets, $\bar{P}_T$ and $\bar{P}_B$ are the corresponding pressures of the droplets, $\epsilon_p$ represents the thickness of the pre-wetted liquid layer away from the droplets, and the droplet profiles $\bar{h}$ are given by the hydrostatic droplet solutions to \eqref{eq:SS}. 
The peak locations $X_T$ and $X_B$ are chosen such that the two droplet profiles meet at a film thickness $h = 0.1$ to initiate the droplet coalescence. 
A typical initial droplet configuration is presented in Figure~\ref{fig:coalesence_angle=90} at time $t^*=0$~ms. Away from the droplets, the precursor thickness $h \equiv \epsilon_p$ and the local flow rate $q \equiv {\epsilon_p^3\Omega}\varphi(\alpha \epsilon_p)/{3}$ form a spatially-uniform steady state solution of the coupled PDE system \eqref{eq:WRM_main}.

We impose the following Dirichlet and Neumann boundary conditions at $x = 0,L$,
\begin{equation}
    h(0,t) = \epsilon_p, \quad q(0,t) = \frac{\epsilon_p^3\Omega}{3}\varphi(\alpha \epsilon_p), \quad h_x(L,t)=0, \quad h_{xx}(L,t) = 0.
\end{equation}
We apply centered finite differences in a Keller box scheme and implicit time-stepping to numerically solve the system \eqref{eq:WRM_main} for the film thickness $h$ and the local flow rate $q$, where the coupled fourth-order PDE system is decomposed into a system of first-order differential equations,
\begin{align}
&\mathcal{K}=h_x, \quad \mathcal{N} = \mathcal{K}_x, \quad \mathcal{G} = q_x, \quad (1+\alpha h)h_t+q_x = 0, \\
&\delta\left(q_t+\Theta_1\frac{q\mathcal{G}}{h}-\Theta_2\frac{q^2\mathcal{K}}{h^2}\right)=
I\left[h\left\{\Omega-\left(\mathcal Z(h)-\mathcal{N}\right)_x\right\}-\frac{3 q}{h^2\varphi(\alpha h)}\right].
\end{align}

\par Figure \ref{fig:coalesence_angle=90} compares the  binary droplet coalescence dynamics on a vertical fibre ($\Omega=1$) predicted by the WRM model \eqref{eq:WRM_main} with the corresponding experimental observations (also shown in Figure~\ref{fig:experiment_90}). In the experiment, the maximum heights $\left(h_{\max}^*\right)$ of the top and bottom droplets are $H^*_T = 0.025$ mm and $H^*_B = 0.05$ mm, respectively, and the fibre radius $R^* = 0.025$ mm. Thus, the larger droplet has approximately twice the height of the smaller droplet.

\par The numerical simulations are performed using the nondimensional parameters and initial configurations corresponding to the experimental conditions.
Specifically, the characteristic thickness is chosen as the maximum height of the larger droplet, $\mathcal{H} = H^*_B = 0.05$ mm, giving the characteristic velocity $\mathcal{U} = g\mathcal{H}^2/\nu = 24.5$ mm/s. The resulting dimensionless parameters are the aspect ratio $\alpha=\mathcal{H}/R^*=2$, $\epsilon= We^{-1/3}=(\rho g \mathcal{H}^2/\sigma)^{1/3}=0.0698$,  the scaling parameter $\eta=\epsilon^2=0.0049$, and the reduced Reynolds number $\delta=\epsilon Re = \epsilon \mathcal{U}\mathcal{H}/\nu = 0.085$. The coating thickness $\epsilon_p^*$ of the thin uniform liquid layer on the prewetted domain cannot be measured directly from experiments. Instead, we choose the dimensionless thickness $\epsilon_p = 0.05$, corresponding to the stabilization parameter $S = 0.047$. The dimensional and dimensionless parameters are summarized in Table \ref{table:nomenclature} in Appendix~\ref{sec:nonmenclature}.

\par After nondimensionalization, the maximum height of the top and bottom droplets are $H_T = 0.5$ and $H_B = 1$, respectively.
The initial configuration of droplets is specified by \eqref{eq:ic}, which describes a small droplet corresponding to pressure $\bar{P}_T=257.89$ centered at $X_T=4.82$ and a large droplet of pressure $\bar{P}_B=208.63$ centered at $X_B=5.18$. The small droplet is obtained numerically by solving the ODE \eqref{eq:SS} over a domain of size $\ell = 0.43$, yielding the corresponding peak height of $H_T=h_{\max} = 0.5$. The large droplet is obtained in the same manner using a domain of size $\ell = 0.5$, with the corresponding peak height of $H_B=h_{\max} = 1$. The model \eqref{eq:WRM_main} is solved on a computational domain of length $L = 10$, which is sufficiently large that the boundaries have no noticeable influence on the coalescence dynamics. To facilitate comparison with the experimental data, the plots in Figure \ref{fig:coalesence_angle=90} are presented using dimensional scales and shifted such that $x^*=0$ coincides with the point where the top droplet meets the precursor layer in the initial configuration.

\par The simulation shows that over the short time interval $0<t^*<0.25$ ms, the droplet profiles relax to form barrel-shaped droplets that rapidly connect, creating a meniscus bridge between them. As early-stage coalescence proceeds, the bridge expands and induces an upward motion towards the smaller droplet, producing a small hump above the larger droplet as it is pulled upward. Over time, this small hump gradually diminishes as the merged droplet continues to migrate upward. The simulation shows overall good agreement with the experimental observation, except for a slightly larger migration distance at time $t^*=1.25$ ms.

\begin{figure}
\centering
\subfloat[]{\includegraphics[height=4.5cm]{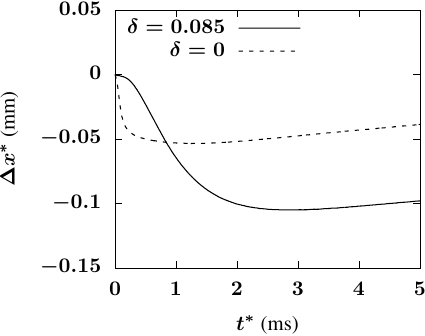}}
\qquad
\subfloat[]{    \includegraphics[height=4.5cm]{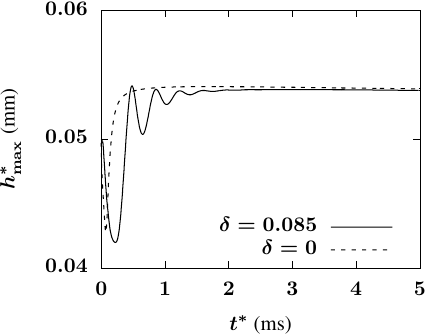}}
    \caption{(a) The center-of-mass shift $\Delta x^*$ and (b) the maximum height of the liquid profile $h^*_{\max}$ over time for cases with and without inertia effects, $\delta=0.085$ (solid curves) and $\delta=0$ (dashed curves). All other system parameters are identical to those used in Figure \ref{fig:coalesence_angle=90}.}
    \label{fig:coalesence_angle=90_cm_hmax_delta_comparison} 
\end{figure}

To quantify the vertical migration of the droplets during coalescence, we trace the center of mass of the droplet pair, $X_c(t)$, defined as
\begin{equation}
    X_c(t) = \dfrac{1}{M}{\bigintsss_{X_T^{\cl}}^{X_B^{\cl}}x\left(h + \frac{\alpha}{2}h^2\right)~\text{d}x},\quad\mbox{where }M={\bigintsss_{X_T^{\cl}}^{X_B^{\cl}} \left(h + \frac{\alpha}{2}h^2\right)~\text{d}x},
\label{eq:COM}
\end{equation}
where $X_T^{\cl}$ and $X_B^{\cl}$ denote the positions of the upper and lower outer contact lines of the droplet pair at time $t$, respectively. The shift in the center of mass relative to the initial configuration is defined as $\Delta x(t) = X_c(t)-X_c(0)$. The solid curve in Figure \ref{fig:coalesence_angle=90_cm_hmax_delta_comparison} (a) shows the dimensional center-of-mass shift $\Delta x^*$ over time. After the merged droplet reaches its maximum upward migration $\Delta x^*=-0.1$ mm at $t^* = 2.8$ ms, gravity causes it to slide down the fibre, resulting in an increase in $\Delta x^*$ for $t^* > 2.8$ ms. 

The inertia effects characterized by the model \eqref{eq:WRM_main} are critical for the migration of the merged droplet over time. To demonstrate the importance of inertia in late-stage coalescence dynamics, we numerically simulate the droplet system without inertia by setting $\delta=0$ in \eqref{eq:WRM_main}, while keeping all other parameters identical to those used in Figure \ref{fig:coalesence_angle=90}.
In this limit, the WRM model \eqref{eq:WRM_main} reduces to a lubrication-based fibre-coating model consistent with previous studies \citep{craster2006viscous,ji2019dynamics} up to a rescaling.
In contrast to the case with moderate inertia effects $(\delta=0.085)$, the simulation shows that the two parent droplets merge rapidly without the apparent secondary hump formation observed in Figure \ref{fig:coalesence_angle=90}. The dashed curve in Figure \ref{fig:coalesence_angle=90_cm_hmax_delta_comparison} (a) shows the center-of-mass shift $\Delta x^*$ for the case without inertia effects ($\delta=0$), indicating that the rapid merging of the two parent droplets induces only a modest upward migration to $\Delta x^* = -0.05$ mm, driven by capillary-dominated early-stage coalescence discussed in Section~\ref{sec:early_coal}.
Subsequently, gravity-dominated dynamics cause the merged droplet to slide down the fibre, with $\Delta x^*$ increasing over time, similar to the case with $\delta=0.085$. The upward migration in the case $\delta=0$ is substantially smaller than that observed in the case with moderate inertia effects $(\delta=0.085)$.

Moreover, the comparison in Figure \ref{fig:coalesence_angle=90_cm_hmax_delta_comparison} (b) 
shows that, when inertia effects are included, the maximum height $h^*_{\max}$ of the liquid profile exhibits oscillations over time as the merged droplet migrates and gradually absorbs the smaller hump above it over several oscillation periods. In contrast, for the inertialess case, after a rapid decay in $h^*_{\max}$ due to relaxation of the initial droplet configuration, $h^*_{\max}$ increases monotonically and approaches a steady maximum value as the merged droplet forms without any oscillations. These results suggest that the inclusion of inertia effects, and the associated formation of secondary humps during asymmetric droplet coalescence, is critical for sustaining the migration of merged droplets against gravity.

\subsection{Hydrodynamic interpretation of the migration mechanism}
\label{sec:migration_hydrodynamic}
Capillary forces associated with the evolving liquid-air interface provide the primary driving mechanism for the droplet migration, while wall friction at the liquid-solid interface provides the dominant viscous resistance. 
In the work of \cite{feng2024short}, these effects were represented using a discrete mass-spring-damper model, where the droplet pair was treated as point masses connected by an effective spring representing surface tension effects, and the viscous friction was modeled by a droplet-size-dependent damper.

\par In the present continuum work, the gravitational, capillary, friction, and nonlinear inertial-transport contributions are obtained directly from the WRM momentum equation \eqref{eq:q_equation}. We define the corresponding depth-integrated force densities by
\begin{align}
    f_g &= I(\alpha h) h\Omega, 
    \quad f_{\text{cap}} = -I(\alpha h)h (\mathcal{Z}(h)-h_{xx})_x, \nonumber\\
    f_{\text{fric}} &= -I(\alpha h)\frac{3q}{h^2\varphi(\alpha h)},
    \quad f_{\text{tr}} = -\delta\left(\Theta_1(\alpha h)\frac{qq_x}{h}-\Theta_2(\alpha h)\frac{q^2h_x}{h^2}\right).
\label{eq:individual_term}
\end{align}
With these definitions, the WRM momentum equation \eqref{eq:q_equation} can be rewritten as
\begin{equation}
    \delta q_t=f_g+f_{\text{cap}}+f_{\text{fric}}+f_{\text{tr}}.
\label{eq:WRM_momentum}
\end{equation}

\par
Next, we relate these force contributions to the motion of the center of mass $X_c$ defined in \eqref{eq:COM}. At the outer contact lines of the droplet pair, the film thickness is equal to the precursor thickness,  $h\left(X_T^{\cl},t\right)=h\left(X_B^{\cl},t\right)=\epsilon_p \ll 1$. We assume that the endpoint fluxes
$q\left(X_T^{\cl},t\right)$ and $q\left(X_B^{\cl},t\right)$ are small compared with the interior contributions. Under these assumptions, the total mass of the droplet pair is approximately conserved, $dM/dt\approx 0$. 
Moreover, integrating the continuity equation in the WRM model \eqref{eq:continuity} over the droplet-pair domain $X_T^{\cl} <x < X_B^{\cl}$, we obtain the approximate migration velocity of the center of mass as
\begin{equation}
    \frac{dX_c}{dt}\approx \frac{1}{M}\int_{X_T^{\cl}}^{X_B^{\cl}}q~\text{d}x.
\label{eq:center_of_mass_time}
\end{equation}
Differentiating \eqref{eq:center_of_mass_time} in time, neglecting the corresponding endpoint corrections, and using \eqref{eq:WRM_momentum}, leads to the estimate of the center-of-mass acceleration,
\begin{equation}
 \delta M \frac{d^2X_c}{dt^2}\approx F_g+F_{\text{cap}}+F_{\text{fric}}+F_{\text{tr}},
 \label{eq:center_of_mass_acceleration}
\end{equation}
where $F_i = \int_{X_T^{\cl}}^{X_B^{\cl}} f_i~\text{d}x$ for $i = \{g, \text{cap},\text{fric}, \text{tr}\}$ denotes the integrated contributions of gravity, capillarity, friction, and inertial transport over the droplet-pair domain. The dimensional force densities and integrated force contributions are related to their dimensionless counterparts by $f_i^* = (\mu \mathcal{U})/\mathcal{H} f_i$ and $F_i^*=(\mu\mathcal{U}\mathcal{L})/\mathcal{H} F_i$, where $i = \{g, \text{cap},\text{fric}, \text{tr}\}$.

\begin{figure}
\centering
\subfloat[]{\includegraphics[height=4cm]{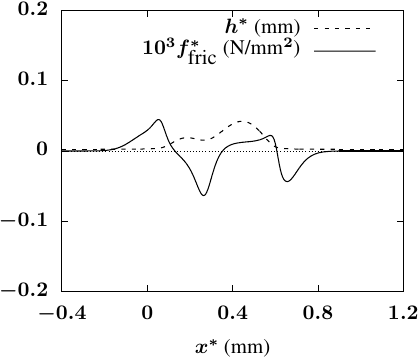}}
\subfloat[]{ \includegraphics[height=4cm]{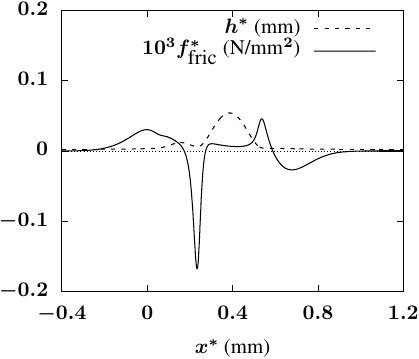}}
\subfloat[]{  \includegraphics[height=4cm]{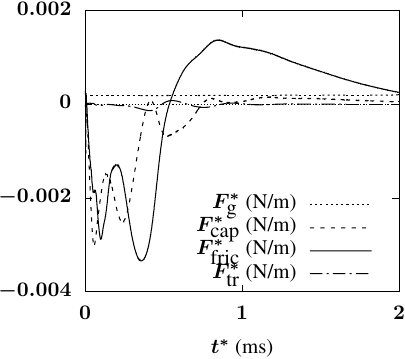}}
    \caption{Friction $f^*_{\text{fric}}$ (solid curves)
    and the corresponding droplet profiles $h^*$ (dashed curves) at (a) $t^*=0.25$ ms and (b) $t^*=0.5$ ms during the coalescence event presented in Figure~\ref{fig:coalesence_angle=90}. (c) Integrated gravitational, capillary, friction, and inertial transport contributions defined in \eqref{eq:center_of_mass_acceleration} to the center-of-mass acceleration over time.
    }
    \label{fig:coalesence_friction}
\end{figure}

\par Equation \eqref{eq:center_of_mass_acceleration} provides a continuum interpretation of the mass-spring-damper-based migration mechanism proposed by \cite{feng2024short}. Instead of representing surface tension and viscous friction effects using springs and dampers with prescribed coefficients, the present hydrodynamic model \eqref{eq:WRM_main} derives the corresponding force distributions consistently from the evolving film thickness and flow rate through \eqref{eq:individual_term}. The capillary contribution $f_{\text{cap}}$ plays the role of an effective spring-like driving force associated with interface curvature and the wetting potential, and $f_{\text{fric}}$ is the continuum counterpart of the damping force. The model also retains the gravitational contribution $f_g$ and the nonlinear inertial-transport contribution $f_{\text{tr}}$, which are not explicitly represented in the simplified discrete model. 
These force distributions emerge naturally from the solutions of the governing equations and provide a quantitative force balance for the migration mechanism proposed qualitatively by \cite{feng2024short}.

\par Figures \ref{fig:coalesence_friction} (a,b) present representative snapshots of the dimensional friction force density $f^*_{\text{fric}}$ and the droplet profiles $h^*$ during the coalescence process shown in Figure~\ref{fig:coalesence_angle=90}. The plots show that the friction at the outer contact line of the larger droplet is predominantly negative, while the friction at the outer contact line of the smaller droplet is positive. In contrast, the friction within the liquid bridge connecting the two droplets remains negative throughout the evolution.
At $t^*=0.25$ ms, the two parent droplets are connected by a relatively thick liquid bridge, and the magnitude of the friction force density within the bridge is comparable to that near the outer contact lines of both droplets.
At $t^*=0.5$ ms, after a secondary hump has formed and the liquid bridge between the hump and the merged droplet becomes thinner,  the magnitude of the wall friction within the liquid bridge becomes significantly larger than that near the outer contact lines of both droplets. Figure \ref{fig:coalesence_friction} (c) shows the dimensionally integrated force contributions, which indicate that the capillary contribution $F^*_{\text{cap}}$ and the friction contribution $F^*_{\text{fric}}$ dominate the center-of-mass dynamics, while gravity and nonlinear inertial transport provide smaller corrections over time. The sum of the four contributions is initially negative, corresponding through \eqref{eq:center_of_mass_acceleration} to an acceleration toward the negative $x$-direction. This is consistent with the observed upward migration of the merged droplet.

\subsection{Effects of parent droplet height ratios and inclination angles}
\label{sec:ratio_inclination}

\par Next, we numerically investigate the effects of the initial height ratio of the parent droplets and the fibre inclination angle on late-stage pairwise droplet coalescence.

\begin{figure}
\centering
\subfloat[]{\includegraphics[height=4.5cm]{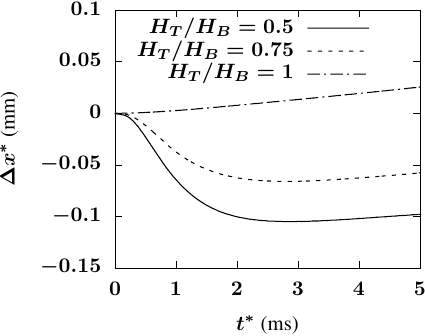}}
\qquad
\subfloat[]{    \includegraphics[height=4.5cm]{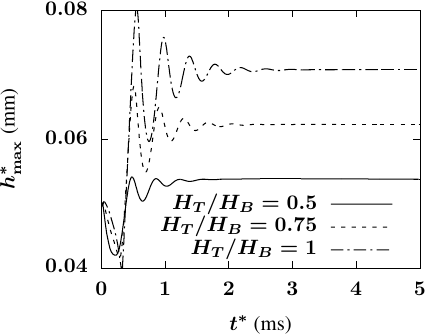}}
    \caption{(a) The shift $\Delta x^*$ in the center of mass of the liquid and (b) the maximum height of the liquid profile $h^*_{\max}$ over time for cases in which the peak-height ratio of the two parent droplets is $H_T/H_B=0.5$ (solid), $H_T/H_B=0.75$ (dashed), and $H_T/H_B=1$ (dot-dashed). All other system parameters are identical to those used in Figure \ref{fig:coalesence_angle=90}.}
    \label{fig:coalesence_angle=90_cm_hmax} 
\end{figure}

We first vary the initial height of the smaller droplet, $H_T$, while keeping the size of the larger droplet, $H_B$, and all other system parameters fixed. Figure \ref{fig:coalesence_angle=90_cm_hmax} (a) shows the center-of-mass shift of the liquid profile over time for different initial parent-droplet height ratios.
A comparison of the dimensional center-of-mass shift $\Delta x^*$ for initial height ratios $H_T/H_B = 0.5$ and $H_T/H_B = 0.75$ shows that increasing the ratio $H_T/H_B$, while keeping the larger droplet size fixed, results in a shorter upward migration distance before gravity-driven sliding down the fibre occurs. When the parent droplets are of the same size $(H_T/H_B = 1)$, symmetric early-stage coalescence is observed, as expected, and the merged droplet does not migrate upward against gravity, leading to a monotonic increase in $\Delta x^*$ over time.
Figure \ref{fig:coalesence_angle=90_cm_hmax} (b) also shows the corresponding evolution of the maximum height $h^*_{\max}$. It indicates that larger height ratios ($H_T/H_B = 0.75$ and $H_T/H_B = 1$) produce larger merged droplets following oscillatory dynamics of similar character but greater magnitude compared to the case with the smaller height ratio $H_T/H_B = 0.5$.

An intuitive explanation for this behavior is that the imbalance in viscous wall friction contributed by the top and bottom droplets, $f_{\text{fric},T}$ and $f_{\text{fric},B}$, depends on the average axial velocities $v_T$ and $v_B$ of the two droplets, where $v_T > 0$ and $v_B < 0$ correspond to the velocities of the top and bottom parent droplets, respectively. 
During the early-stage coalescence, the condition $|v_T| > |v_B|$ arises from the difference in Laplace pressures of the two droplets, with $\bar{P}_T > \bar{P}_B$, which was verified experimentally in \cite{feng2024short}. 
Under the assumption that the friction $f_{\text{fric}}$ is proportional to the droplet velocity \citep{lorenceau2004drops}, i.e., $|f_{\text{fric}}| \propto |v|$, it follows that $|f_{\text{fric}, T}| > |f_{\text{fric},B}|$. Since $f_{\text{fric}, T}$ acts in the negative $x$ direction and $f_{\text{fric}, B}$ acts in the positive $x$ direction, the net wall friction points in the negative $x$ direction.

Moreover, during binary droplet coalescence, the magnitude of each parent droplet's velocity is inversely proportional to its mass \citep{feng2024short}, which leads to the relation $|v_T|/|v_B| \approx [(H_B + R) / (H_T + R)]^3$, where $R$ is the dimensionless radius of the fibre. Consequently, for a fixed $H_B$ (the height of the larger droplet), increasing $H_T$ (the height of the smaller droplet) decreases the velocity ratio $|v_T|/|v_B|$ and makes the motions of the two droplets more symmetric. At the same time, the Laplace-pressure asymmetry decreases as the droplet sizes become more similar. 
The reduced asymmetry in both the capillary and viscous friction contributions weakens the net force imbalance responsible for upward migration, resulting in a slower upward migration of the center of mass (see the comparison between the $H_T/H_B = 0.75$ and $H_T/H_B=0.5$ cases in Figure~\ref{fig:coalesence_angle=90_cm_hmax} (a)).

\par
\begin{figure}
\centering
\includegraphics[height=4.5cm]{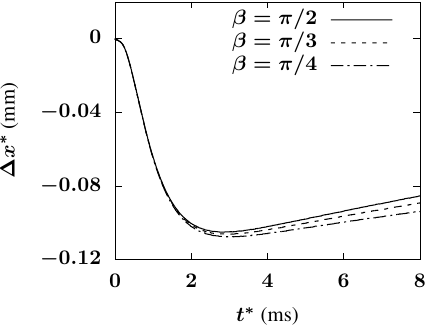}
    \caption{The center-of-mass shift $\Delta x^*$ for inclination angles $\beta=\pi/2$ (solid), $\beta=\pi/3$ (dashed), and $\beta=\pi/4$ (dot-dashed).
   All other settings are identical to those in Figure~\ref{fig:coalesence_angle=90}.
    }
    \label{fig:coalesence_angle}
\end{figure}

 The inclination angle $\beta$ of the fibre controls the relative importance of gravity and other competing physical effects in the droplet-fibre system. In Figure \ref{fig:coalesence_angle}, we plot the center-of-mass shift $\Delta x^*$ from binary droplet coalescence simulations using the same settings as in Figure \ref{fig:coalesence_angle=90}, while varying the inclination angle $\beta$ with $\beta = \pi/2$, $\pi/3$, and $\pi/4$. The results indicate that the early- to late-stage coalescence of the two parent droplets is not strongly dependent on the inclination angle, as $\Delta x^*$ follows nearly identical trajectories in time until reaching the maximum upward position. In the final stage, after the merged droplet has stabilized, gravity becomes dominant and the inclination angle affects the sliding speed of the droplet down the fibre. Specifically, for smaller inclination angles $\beta$, the gravity parameter $\Omega$ is reduced, and the merged droplet slides down the fibre more slowly as expected. In the limit $\beta \to 0$, the system reduces to the horizontal fibre case discussed in \cite{feng2024short}, in which the merged droplet is expected to become stationary after coalescence is completed, in the absence of gravitational effects.

\section{Conclusions and discussions}
\label{sec:discussion}
This paper focused on the study of droplet coalescence events on a pre-wetted sloped cylinder, where the merged droplet migrates towards the smaller droplet. For early-stage coalescence, we used a lubrication model to capture the influence of the imbalanced Laplace pressure on the parent droplets. Through asymptotic analysis, we showed that the liquid bridge connecting two parent droplets migrates toward the smaller droplet in a self-similar manner during coalescence. This prediction is consistent with our experimental observations in terms of both liquid bridge profiles and measured migration velocity. For the late-stage coalescence dynamics, we developed a weighted residual boundary layer model to account for the impact of moderate inertia, gravity, surface tension, and intermolecular forces on the dynamics. Using this model, we numerically demonstrated the effects of inertia, droplet size ratio, inclination angle, viscous friction, and surface tension on the shift in the center of mass of the liquid profile.

In the current work, we assumed a two-dimensional axisymmetric barrel-shaped droplet to simplify the modeling and numerical computation, and we limited the model to account for only low to moderate inertia effects. More interesting results are expected if some of these assumptions are relaxed, such as considering non-axisymmetric droplet configurations caused by gravity in the normal direction or the dynamics of droplet coalescence in a three-dimensional setting. Specifically, the transition from axisymmetric barrel-shaped droplets to non-axisymmetric clamshell-shaped ones has been analyzed and experimentally observed \citep{gupta2021effect}. Non-axisymmetric droplet coalescence on a cylindrical domain influenced by gravity can lead to richer dynamics.
Furthermore, it is also of interest to thoroughly investigate the impact of high inertia effects, which naturally occur for large droplets on a thin wire and can induce more complex instabilities such as rocking modes \citep{poulain2023sliding}. Another interesting topic is the coalescence of multiple droplets on a sloped cylindrical surface, where a cascade of merged-droplet migration can be achieved. Such a study has potential applications across a broad range of fields, including microfluidics, filtration, and fog collection, which involve droplet coalescence on various geometries. Moreover, the transition between instantaneous coalescence and temporary non-coalescence of sessile droplets of different yet completely miscible liquids on a flat substrate has been extensively studied by \cite{karpitschka2014coalescence} and \cite{Karpitschka_Riegler_2014}. Extending the present study to the coalescence dynamics of droplets from different miscible liquids on fibres or other non-planar substrate geometries is expected to reveal similarly rich interfacial transport phenomena.

\section*{Acknowledgments}
H.J. acknowledges support from the National Science Foundation (NSF) under Grant No. DMS-2309774.

\section*{Declaration of Interests}
The authors report no conflict of interest.

\section*{Use of Artificial Intelligence (AI) Tools}
The authors used ChatGPT solely for language editing, including grammar correction and stylistic refinement of portions of the manuscript. All scientific content, analyses, figures, interpretations, and conclusions were developed and verified by the authors.

\appendix

\section{The expressions of $\chi(\zeta)$, $\psi(\zeta)$ and $\Upsilon(\zeta)$ of equation \eqref{eq:WRM_main}}
\label{appendix:a}
\begin{equation*}
\chi(\zeta)=130-441(1+\zeta)^2+622(1+\zeta)^4-301(1+\zeta)^6+16\text{ln}(1+\zeta)
\end{equation*}
\begin{equation*}
+4(1+\zeta)^2\text{ln}(1+\zeta)\left\{197(1+\zeta)^4-234(1+\zeta)^2+78\right.
\end{equation*}
\begin{equation*}
\left.+6\text{ln}(1+\zeta)\left[16(1+\zeta)^4\text{ln}(1+\zeta)-36(1+\zeta)^4+22(1+\zeta)^2+3\right]\right\},
\end{equation*}

\begin{multline*}
 \psi(\zeta)=17(1+\zeta)^6-30(1+\zeta)^4+15(1+\zeta)^2\\
 +12(1+\zeta)^4\text{ln}(1+\zeta)\left[2(1+\zeta)^2\text{ln}(1+\zeta)-3(1+\zeta)^2+2\right]-2,
\end{multline*}
    
\begin{equation*}
\Upsilon(\zeta)=9\left[4(1+\zeta)^2\ln(1+\zeta)\left[9+6(1+\zeta)^2\text{ln}(1+\zeta)\left\{4(1+\zeta)^2\text{ln}(1+\zeta)\left(4(1+\zeta)^4\text{ln}(1+\zeta)\right.\right.\right.\right.
\end{equation*}
\begin{equation*}
\left.\left.-12(1+\zeta)^4+7(1+\zeta)^2+2\right)+61(1+\zeta)^6-69(1+\zeta)^4+9(1+\zeta)^2+9\right\}
\end{equation*}
\begin{equation*}
\left.+(1+\zeta)^2\left\{58-303(1+\zeta)^2+456(1+\zeta)^4-220(1+\zeta)^6\right\}\right]
\end{equation*}
\begin{equation*}
\left.+\zeta^2(2+\zeta)^2\left(153(1+\zeta)^6-145(1+\zeta)^4+53(1+\zeta)^2-1\right)\right](1+\zeta).
\end{equation*}

\section{Nomenclature}
\label{sec:nonmenclature}
Table \ref{table:nomenclature} summarizes the nomenclature and corresponding parameter values for a typical binary droplet coalescence experiment on a fibre. The upper portion of the table lists dimensional parameters together with their units, and the lower portion lists the corresponding nondimensional parameters used in the simulations.

\begin{table}
\centering
\begin{tabular} {lcccc}
 \textbf{Definition}&& \textbf{Symbol}&& \textbf{Sample Values}\\
 &&&& \\
 Radius of the fibre (mm)&& $R^*$ && 0.025  \\
maximum height of droplets (mm) && $h_{\max}^*$  && $0.025 - 0.05$  \\

 lengthscale in radial direction (mm)  && $\mathcal{H}$  && 0.05 \\
 lengthscale in streamwise direction (mm) && $\mathcal{L}$ && 0.72 \\
characteristic streamwise velocity (mm/s)  &&$\mathcal{U}$ && 24.5
\\
&&&&\\

aspect ratio $\mathcal{H}/R^*$  && $\alpha$ && 2 \\
scaling parameter $(\mathcal{H}/\mathcal{L})^2$  &&$\eta$  && 0.0049  \\
uniform layer thickness  &&$\epsilon_p$  && 0.05 \\
stabilization parameter   &&$S$  && 0.047    \\
reduced Reynolds number  && $\delta$  && 0.085  \\
streamwise gravity parameter $\sin\beta$ && $\Omega$ && $0.5 - 1$\\
maximum droplet height && $H_T$, $H_B$ && $0.5 -1$
 \end{tabular}
  \caption{Nomenclature and their sample values.}
  \label{table:nomenclature}
\end{table}

\bibliographystyle{jfm}
\bibliography{dropCoalescence}

@preamble{ " \newcommand{\noop}[1]{} " }

@article{bharti2023plateau,
  title={{Plateau-Rayleigh} instability of a viscous film on a soft fiber},
  author={Bharti, Bharti and Carlson, Andreas and Chan, Tak Shing and Salez, Thomas},
  journal={arXiv preprint arXiv:2312.11962},
  year={2023}
}

@article{chattopadhyay2024modeling,
  title={Modeling reactive film flows down a heated fiber},
  author={Chattopadhyay, Souradip and Ji, Hangjie},
  journal={Chemical Engineering Science},
  volume={300},
  pages={120551},
  year={2024},
  publisher={Elsevier}
}

@article{karpitschka2014coalescence,
  title={Coalescence and noncoalescence of sessile drops: {I}mpact of surface forces},
  author={Karpitschka, Stefan and Hanske, Christoph and Fery, Andreas and Riegler, Hans},
  journal={Langmuir},
  volume={30},
  number={23},
  pages={6826--6830},
  year={2014},
  publisher={ACS Publications}
}

@article{gupta2021effect,
  title={Effect of gravity on the shape of a droplet on a fiber: {N}early axisymmetric profiles with experimental validation},
  author={Gupta, Ankur and Konicek, Andrew R and King, Mark A and Iqtidar, Azmaine and Yeganeh, Mohsen S and Stone, Howard A},
  journal={Physical Review Fluids},
  volume={6},
  number={6},
  pages={063602},
  year={2021},
  publisher={APS}
}

@article{pawar2019symmetric,
  title={Symmetric and asymmetric coalescence of droplets on a solid surface in the inertia-dominated regime},
  author={Pawar, Nilesh D and Bahga, Supreet Singh and Kale, Sunil R and Kondaraju, Sasidhar},
  journal={Physics of Fluids},
  volume={31},
  number={9},
  year={2019},
  publisher={AIP Publishing}
}

@article{bonn2009wetting,
  title={Wetting and spreading},
  author={Bonn, Daniel and Eggers, Jens and Indekeu, Joseph and Meunier, Jacques and Rolley, Etienne},
  journal={Reviews of Modern Physics},
  volume={81},
  number={2},
  pages={739--805},
  year={2009},
  publisher={APS}
}

@article{christianto2022modeling,
  title={Modeling the dynamics of partially wetting droplets on fibers},
  author={Christianto, Raymond and Rahmawan, Yudi and Semprebon, Ciro and Kusumaatmaja, Halim},
  journal={Physical Review Fluids},
  volume={7},
  number={10},
  pages={103606},
  year={2022},
  publisher={APS}
}

@article{li2013fastest,
  title={The fastest drop climbing on a wet conical fibre},
  author={Li, Er Qiang and Thoroddsen, Sigurdur T},
  journal={Physics of Fluids},
  volume={25},
  number={5},
  year={2013},
  publisher={AIP Publishing}
}

@article{jiang2022coalescence,
  title={Coalescence-induced propulsion of droplets on a superhydrophilic wire},
  author={Jiang, Youhua and Feng, Leyun and O'Donnell, Allison and Machado, Christian and Choi, Wonjae and Patankar, Neelesh A and Park, Kyoo-Chul},
  journal={Applied Physics Letters},
  volume={121},
  number={23},
  year={2022},
  publisher={AIP Publishing}
}

@article{hernandez2012symmetric,
  title={Symmetric and asymmetric coalescence of drops on a substrate},
  author={Hern{\'a}ndez-S{\'a}nchez, JF and Lubbers, LA and Eddi, Antonin and Snoeijer, JH},
  journal={Physical Review Letters},
  volume={109},
  number={18},
  pages={184502},
  year={2012},
  publisher={APS}
}

@article{Karpitschka_Riegler_2014, 
title={Sharp transition between coalescence and non-coalescence of sessile drops}, volume={743}, 
DOI={10.1017/jfm.2014.73}, 
journal={Journal of Fluid Mechanics}, 
author={Karpitschka, Stefan and Riegler, Hans}, 
year={2014}, 
pages={R1}}

@article{dekker2022elasticity,
  title={When elasticity affects drop coalescence},
  author={Dekker, Pim J and Hack, Michiel A and Tewes, Walter and Datt, Charu and Bouillant, Ambre and Snoeijer, Jacco H},
  journal={Physical Review Letters},
  volume={128},
  number={2},
  pages={028004},
  year={2022},
  publisher={APS}
}

@article{oron1997long,
  title={Long-scale evolution of thin liquid films},
  author={Oron, Alexander and Davis, Stephen H and Bankoff, S George},
  journal={Reviews of modern physics},
  volume={69},
  number={3},
  pages={931},
  year={1997},
  publisher={APS}
}

@article{eggers2024coalescence,
  title={Coalescence Dynamics},
  author={Eggers, Jens and Sprittles, James E and Snoeijer, Jacco H},
  journal={Annual Review of Fluid Mechanics},
  volume={57},
  year={2024},
  publisher={Annual Reviews}
}

@article{fournier2021droplet,
  title={Droplet migration on conical fibers},
  author={Fournier, Clementine and Lee, Carmen L and Schulman, Rafael D and Rapha{\"e}l, {\'E}lie and Dalnoki-Veress, Kari},
  journal={The European Physical Journal E},
  volume={44},
  pages={1--6},
  year={2021},
  publisher={Springer}
}

@article{lorenceau2004drops,
  title={Drops on a conical wire},
  author={Lorenceau, Elise and Qu{\'e}r{\'e}, David},
  journal={Journal of Fluid Mechanics},
  volume={510},
  pages={29--45},
  year={2004},
  publisher={Cambridge University Press}
}

@article{chan2021film,
  title={Film coating by directional droplet spreading on fibers},
  author={Chan, Tak Shing and Lee, Carmen L and Pedersen, Christian and Dalnoki-Veress, Kari and Carlson, Andreas},
  journal={Physical Review Fluids},
  volume={6},
  number={1},
  pages={014004},
  year={2021},
  publisher={APS}
}

@article{jiang2019fog,
  title={Fog collection on a superhydrophilic wire},
  author={Jiang, Youhua and Savarirayan, Shaan and Yao, Yuehan and Park, Kyoo-Chul},
  journal={Applied Physics Letters},
  volume={114},
  number={8},
  year={2019},
  publisher={AIP Publishing}
}

@article{sadeghpour2019water,
  title={Water vapor capturing using an array of traveling liquid beads for desalination and water treatment},
  author={Sadeghpour, A and Zeng, Z and Ji, H and Dehdari Ebrahimi, N and Bertozzi, AL and Ju, YS},
  journal={Science Advances},
  volume={5},
  number={4},
  pages={eaav7662},
  year={2019},
  publisher={American Association for the Advancement of Science}
}

@article{brunazzi2000design,
  title={Design of complex wire-mesh mist eliminators},
  author={Brunazzi, Elisabetta and Paglianti, Alessandro},
  journal={AIChE Journal},
  volume={46},
  number={6},
  pages={1131--1137},
  year={2000},
  publisher={Wiley Online Library}
}

@article{kowalski2022dynamics,
  title={Dynamics of fog droplets on a harp wire},
  author={Kowalski, Nicholas G and Boreyko, Jonathan B},
  journal={Soft Matter},
  volume={18},
  number={37},
  pages={7148--7158},
  year={2022},
  publisher={Royal Society of Chemistry}
}

@article{zhao2023slip,
  title={Slip-enhanced {Rayleigh--Plateau} instability of a liquid film on a fibre},
  author={Zhao, Chengxi and Zhang, Yixin and Si, Ting},
  journal={Journal of Fluid Mechanics},
  volume={954},
  pages={A46},
  year={2023},
  publisher={Cambridge University Press}
}

@article{chan2020directional,
  title={Directional spreading of a viscous droplet on a conical fibre},
  author={Chan, Tak Shing and Yang, Fan and Carlson, Andreas},
  journal={Journal of Fluid Mechanics},
  volume={894},
  pages={A26},
  year={2020},
  publisher={Cambridge University Press}
}

@article{carroll1976accurate,
  title={The accurate measurement of contact angle, phase contact areas, drop volume, and {Laplace} excess pressure in drop-on-fiber systems},
  author={Carroll, BJ},
  journal={Journal of Colloid and Interface Science},
  volume={57},
  number={3},
  pages={488--495},
  year={1976},
  publisher={Elsevier}
}

@article{feng2024short,
  title={Short-time asymmetric droplet coalescence dynamics on a pre-wetted fiber},
  author={Feng, Leyun and Jiang, Youhua and Machado, Christian and Choi, Wonjae and Patankar, Neelesh A and Park, Kyoo-Chul},
  journal={Applied Physics Letters},
  volume={125},
  number={6},
  year={2024},
  publisher={AIP Publishing}
}

@article{ji2021thermally,
  title={Thermally-driven coalescence in thin liquid film flowing down a fibre},
  author={Ji, Hangjie and Falcon, Claudia and Sedighi, Erfan and Sadeghpour, Abolfazl and Ju, Y Sungtaek and Bertozzi, Andrea L},
  journal={Journal of Fluid Mechanics},
  volume={916},
  pages={A19},
  year={2021},
  publisher={Cambridge University Press}
}

@article{sadeghpour2021experimental,
  title={Experimental study of a string-based counterflow wet electrostatic precipitator for collection of fine and ultrafine particles},
  author={Sadeghpour, A. and Oroumiyeh, F. and Zhu, Y. and Ko, D. D. and Ji, H. and Bertozzi, A. L. and Ju, Y. S.},
  journal={Journal of the Air \& Waste Management Association},
  pages={1--15},
  year={2021},
  publisher={Taylor \& Francis}
}

@article{ji2019dynamics,
  title={Dynamics of thin liquid films on vertical cylindrical fibres},
  author={Ji, H. and Falcon, C. and Sadeghpour, A. and Zeng, Z. and Ju, Y. S. and Bertozzi, A. L.},
  journal={Journal of Fluid Mechanics},
  volume={865},
  pages={303--327},
  year={2019},
  publisher={Cambridge University Press}
}

@article{glasner2003coarsening,
  title={Coarsening dynamics of dewetting films},
  author={Glasner, Karl B and Witelski, Thomas P},
  journal={Physical review E},
  volume={67},
  number={1},
  pages={016302},
  year={2003},
  publisher={APS}
}

@article{zhao2024inertia,
  title={Inertia and slip effects on the instability of a liquid film coated on a fibre},
  author={Zhao, Chengxi and Qiao, Ran and Mu, Kai and Si, Ting and Luo, Xisheng},
  journal={Journal of Fluid Mechanics},
  volume={982},
  pages={A13},
  year={2024},
  publisher={Cambridge University Press}
}

@article{ashgriz1990coalescence,
  title={Coalescence and separation in binary collisions of liquid drops},
  author={Ashgriz, N and Poo, JY},
  journal={Journal of Fluid Mechanics},
  volume={221},
  pages={183--204},
  year={1990},
  publisher={Cambridge University Press}
}

@article{eggers1999coalescence,
  title={Coalescence of liquid drops},
  author={Eggers, Jens and Lister, John R and Stone, Howard A},
  journal={Journal of Fluid Mechanics},
  volume={401},
  pages={293--310},
  year={1999},
  publisher={Cambridge University Press}
}

@article{sprittles2014dynamics,
  title={Dynamics of liquid drops coalescing in the inertial regime},
  author={Sprittles, James E and Shikhmurzaev, Yulii D},
  journal={Physical Review E},
  volume={89},
  number={6},
  pages={063008},
  year={2014},
  publisher={APS}
}

@article{kaneelil2025coalescence,
  title={Coalescence of viscoelastic sessile drops: the small and large contact angle limits},
  author={Kaneelil, Paul R and Tojo, Kazuki and Farsoiya, Palas Kumar and Deike, Luc and Stone, Howard A},
  journal={arXiv preprint arXiv:2505.02226},
  year={2025}
}

@article{gilet2009digital,
  title={Digital microfluidics on a wire},
  author={Gilet, Tristan and Terwagne, Denis and Vandewalle, Nicolas},
  journal={Applied Physics Letters},
  volume={95},
  number={1},
  year={2009},
  publisher={AIP Publishing}
}

@article{poulain2023sliding,
  title={Sliding, vibrating and swinging droplets on an oscillating fibre},
  author={Poulain, St{\'e}phane and Carlson, Andreas},
  journal={Journal of Fluid Mechanics},
  volume={967},
  pages={A24},
  year={2023}
}

@article{reisfeld1992non,
  title={Non-isothermal flow of a liquid film on a horizontal cylinder},
  author={Reisfeld, B and Bankoff, SG},
  journal={Journal of Fluid Mechanics},
  volume={236},
  pages={167--196},
  year={1992},
  publisher={Cambridge University Press}
}

@article{RevModPhys.57.827,
  title = {Wetting: statics and dynamics},
  author = {de Gennes, P. G.},
  journal = {Rev. Mod. Phys.},
  volume = {57},
  issue = {3},
  pages = {827--863},
  numpages = {0},
  year = {1985},
  month = {Jul},
  publisher = {American Physical Society},
  doi = {10.1103/RevModPhys.57.827},
  url = {https://link.aps.org/doi/10.1103/RevModPhys.57.827}
}

@book{kalliadasis2011falling,
  title={Falling liquid films},
  author={Kalliadasis, Serafim and Ruyer-Quil, Christian and Scheid, Benoit and Velarde, Manuel Garc{\'\i}a},
  volume={176},
  year={2011},
  publisher={Springer Science \& Business Media}
}

@article{haefner2015influence,
  title={Influence of slip on the {P}lateau--{R}ayleigh instability on a fibre},
  author={Haefner, Sabrina and Benzaquen, Michael and B{\"a}umchen, Oliver and Salez, Thomas and Peters, Robert and McGraw, Joshua D and Jacobs, Karin and Rapha{\"e}l, Elie and Dalnoki-Veress, Kari},
  journal={Nature communications},
  volume={6},
  pages={7409},
  year={2015},
  publisher={Nature Publishing Group}
}

@article{duprat2007absolute,
  title={Absolute and convective instabilities of a viscous film flowing down a vertical fiber},
  author={Duprat, C and Ruyer-Quil, C and Kalliadasis, S and Giorgiutti-Dauphin{\'e}, F},
  journal={Physical Review Letters},
  volume={98},
  number={24},
  pages={244502},
  year={2007},
  publisher={APS}
}

@article{craster2006viscous,
  title={On viscous beads flowing down a vertical fibre},
  author={Craster, R. V. and Matar, O. K.},
  journal={Journal of Fluid Mechanics},
  volume={553},
  pages={85--105},
  year={2006},
  publisher={Cambridge University Press}
}

@article{ruyer2008modelling,
  title={Modelling film flows down a fibre},
  author={Ruyer-Quil, C and Treveleyan, P and Giorgiutti-Dauphin{\'e}, F and Duprat, C and Kalliadasis, S},
  journal={Journal of Fluid Mechanics},
  volume={603},
  pages={431--462},
  year={2008},
  publisher={Cambridge University Press}
}

@article{novbari2018parametric,
  title={Parametric excitation of an axisymmetric flow of a thin liquid film down a vertical fiber},
  author={Novbari, E and Oron, A},
  journal={Acta Mechanica},
  volume={229},
  number={2},
  pages={549--569},
  year={2018},
  publisher={Springer}
}
\end{document}